\documentclass[twocolumn]{aastex61}

\usepackage[utf8]{inputenc}
\usepackage[T5,T1]{fontenc}

\received{}
\revised{}
\accepted{}
\submitjournal{ApJ}

\shorttitle{Two different grain size distributions}
\shortauthors{Ohashi et al.}

\begin{document}

\title{Two Different Grain Size Distributions within the Protoplanetary Disk around HD 142527 Revealed by ALMA Polarization Observation}

\author{Satoshi Ohashi}
\affiliation{RIKEN Cluster for Pioneering Research, 2-1, Hirosawa, Wako-shi, Saitama 351-0198, Japan}

\author{Akimasa Kataoka}
\affiliation{National Astronomical Observatory of Japan, 2-21-1 Osawa, Mitaka, Tokyo 181-8588, Japan}

\author{Hiroshi Nagai}
\affiliation{National Astronomical Observatory of Japan, 2-21-1 Osawa, Mitaka, Tokyo 181-8588, Japan}

\author{Munetake Momose}
\affiliation{College of Science, Ibaraki University, 2-1-1 Bunkyo, Mito, Ibaraki 310-8512, Japan}

\author{Takayuki Muto}
\affiliation{Division of Liberal Arts, Kogakuin University, 1-24-2 Nishi-Shinjuku, Shinjuku-ku, Tokyo 163-8677, Japan}

\author{Tomoyuki Hanawa}
\affiliation{Center for Frontier Science, Chiba University, 1-33 Yayoi-cho, Inage, Chiba 263-8522, Japan}

\author{Misato Fukagawa}
\affiliation{Division of Particle and Astrophysical Science, Graduate School of Science, Nagoya University, Furo-cho, Chikusa-ku, Nagoya, Aichi 464-8602, Japan}

\author{Takashi Tsukagoshi}
\affiliation{National Astronomical Observatory of Japan, 2-21-1 Osawa, Mitaka, Tokyo 181-8588, Japan}
\affiliation{College of Science, Ibaraki University, 2-1-1 Bunkyo, Mito, Ibaraki 310-8512, Japan}

\author{Kohji Murakawa}
\affiliation{College of General Education, Osaka Sangyo University, 3-1-1, Nakagaito, Daito, Osaka 574-8530, Japan}

\author{Hiroshi Shibai}
\affiliation{Graduate School of Science, Osaka University, 1-1 Machikaneyama, Toyonaka, Osaka 560-0043, Japan}



\begin{abstract}
The origin of polarized emission from protoplanetary disks is uncertain. Three mechanisms have been proposed for such polarized emission: grain alignment with magnetic fields, grain alignment with radiation gradients, and self-scattering of thermal dust emission. Aiming to observationally identify the polarization mechanisms, we present ALMA polarization observations of the 0.87 mm dust continuum emission toward the circumstellar disk around HD 142527 with high spatial resolution. 
We confirm that the polarization vectors in the northern region are consistent with self-scattering. Furthermore, we show that the polarization vectors in the southern region are consistent with grain alignment by magnetic fields, although self-scattering cannot be ruled out.
To understand the differences between the polarization mechanisms, we propose a simple grain size segregation model: small dust grains ($\lesssim$ 100 microns) are dominant and aligned with magnetic fields in the southern region, and middle-sized ($\sim100$ microns) grains in the upper layer emit self-scattered polarized emission in the northern region. The grain size near the middle plane in the northern region cannot be measured because the emission at 0.87 mm is optically thick. However, it can be speculated that larger dust grains ($\gtrsim$ cm) may accumulate near this plane.
These results are consistent with those of a previous analysis of the disk, in which large grain accumulation and optically thick emission from the northern region were found. This model is also consistent with theories where smaller dust grains are aligned with magnetic fields. The magnetic fields are toroidal, at least in the southern region.

\end{abstract}




\section{Introduction} \label{sec:intro}
In general, magnetic fields are believed to play an important role in grain alignment, which causes polarized dust emission in the interstellar medium (ISM) and star-forming regions such as molecular clouds and dense cores \citep[see review by][]{laz03}.
The processes of grain alignment with magnetic fields have been studied since the discovery of polarization by \citet{hal49,hin49}. 
\citet{dra96,dra97} showed that grains are aligned by radiation torques (RATs) with respect to the magnetic fields under anisotropic radiation fields, indicating that in addition to magnetic fields, RATs are important in the alignment of grains.
As a result of the grain alignment, polarized emission can be observed with the polarization vectors perpendicular to the direction of the magnetic fields \citep{laz07,and15}.
Therefore, dust polarization observations have been used to investigate the magnetic fields from molecular clouds to dense cores, protostars, and protostellar disks in various star forming regions \citep[e.g.,][]{rao98,lai01,gir06,rao14,cox15,seg15,hul13,hul14,hul17}.

In contrast, the polarization mechanisms in protoplanetary disks are under debate. Three mechanisms have been proposed for the polarization of millimeter and submillimeter emission. They are different from those in the ISM since the size and opacity of dust grains in such disks are different from those in the ISM.

The two possible origins of dust polarization at millimeter and submillimeter wavelengths, namely magnetic grain alignment and radiative grain alignment, are related to the thermal emission of aligned dichroic grains \citep{cho07,taz17}. The long axis of dust grains tends to be perpendicular to magnetic fields in the former case, whereas it tends to be perpendicular to radiation gradients in the latter case. These two alignment processes compete with each other.

The first mechanism is grain alignment with magnetic fields. \citet{cho07} performed synthetic polarization observations of protoplanetary disks assuming that dust grains are aligned with toroidal magnetic fields. They found that the polarization vectors point in radial directions due to the grain alignment with magnetic fields.
Toroidal magnetic fields are thought to be amplified by magnetorotational instability  (MRI) and are reproduced in numerical simulations of rotating disks \citep[e.g.,][]{bra95,fro06,hen09,dav10,bai13,suz14}.

The second mechanism is grain alignment with radiation gradients.
\citet{lazho07} found that millimeter-sized grains are more likely to be aligned with the direction of the radiation fields than with that of the magnetic fields due to their relatively slow Larmor precession in the latter.
\citet{taz17} performed radiative transfer calculations of millimeter-wave polarization in protoplanetary disks by applying RAT theory, and found that grains larger than a few 10 $\mu$m are aligned with their short axis in the direction of the radiation fields rather than that of the magnetic fields. 
Therefore, the polarization vectors trace the direction of radiation anisotropy.

The third mechanism is the self-scattering theory proposed by \citet{kat15}.
If dust grains are smaller than the observation wavelength, the scattering opacity is low but the polarization degree can be high, as expected in the Rayleigh scattering regime. In contrast, if dust grains are larger than the observation wavelength, the scattering opacity becomes high but the polarization degree is low. 
Therefore, the continuum emission is expected to be polarized appreciably only when the grain size is comparable to the observation wavelength ($\lambda \sim 2\pi a_{\rm max}$, where $\lambda$ is the observation wavelength and $a_{\rm max}$ is the maximum grain size) and when the radiation is anisotropic.
The importance of dust scattering was addressed by \cite{yan16a,yan16b,poh16,kat16a}.

We note that the grain alignment by mechanical torques may also be important as a fourth mechanism. The mechanical alignment occurs in dense regions where gas dynamics is more important than radiation  \citep[e.g.,][]{gol52a,gol52b}.  Recent studies have shown the importance of mechanical grain alignment \citep{hoa18}. However, we focus here on the former three mechanisms (grain alignment with the magnetic fields, grain alignment with the radiation gradients, and self-scattering) which have been previously discussed for protoplanetary disks.

For observations, \citet{hug09,hug13} tried to detect dust polarization from circumstellar disks using millimeter and submillimeter arrays, namely the Combined Array for Research in Millimeter-wave (CARMA) and the Submillimeter Array (SMA). However, they obtained an upper limit of only $\simeq1$\%.
The first detection of polarization from T Tauri disks was performed by \citet{ste14}, who observed 1.25 mm polarized emission from the HL Tau disk using CARMA and SMA, and found that the polarization vectors are aligned with the minor axis of the disk. They interpreted these results as indicating that the magnetic fields are coincident with the major axis of the disk, which is more consistent with toroidal fields than with poloidal fields, although it is inconsistent with simple toroidal magnetic fields. However, \citet{kat16a,yan16a} have indicated that this polarization morphology can be explained by self-scattering rather than grain alignment with magnetic fields.
Indeed, polarized emission from some other disks is likely from self-scattering rather than grain alignment with magnetic fields \citep[e.g.,][]{fer16,lee18,cox18,gir18,hul18,sad18,har18}.

Recent Atacama Large Millimeter/submillimeter Array (ALMA) observations of HL Tau show a dependence of the polarization on wavelength \citep[][]{kat17,ste17}.
The polarization vectors lie in the azimuthal direction at a 3 mm wavelength, which is consistent with the radiative grain alignment theory, whereas they lie in the minor axis direction at a 0.87 mm wavelength, supporting self-scattering.
These results indicate that the polarization mechanism depends on the observation wavelength.

In this study, we report new ALMA polarization observations toward the circumstellar disk around HD 142527 at 0.87 mm. The new observations have an angular resolution that is almost twice that used by Kataoka et al. (2016b, henceforth \citet{kat16}), who discovered a flip of the polarization, i.e., a change in the polarization vector directions from radial to azimuthal, around the disk outer edge.
This flip of polarization vectors is consistent with the morphology expected with self-scattering due to a change in the direction of the thermal dust radiation flux.
However, \citet{taz17} pointed out that the flip of the polarization vectors can also be explained by radiative grain alignment caused by a change in the direction of the radiative flux.
In addition, the polarization vectors in the southern region point only in radial directions, which may be inconsistent with the polarization expected with self-scattering.
Therefore, we aim to clarify the polarization mechanisms in the entire disk with the new higher spatial resolution data.

HD 142527 is a Herbig Ae star and is known to have a transitional disk, which has a cavity of dust emission \citep{str89,cal02,and11}.
The stellar mass was evaluated to be 2.2 $M_\odot$ based on the evolutionary track on the Hertzsprung-Russell diagram \citep{ver11}.
A companion with a mass of $0.1-0.4$ $M_\odot$ was found at about $\sim15$ au from the central star \citep{bil12,clo14,rod14,lac16}.
Here, we adopt the distance measured by {\it Gaia}, $156\pm7.5$ pc \citep{gai16a,gai16b}.

Using ALMA high spatial resolution observations, \citet{cas13,fuk13} showed that the transitional disk of HD 142527 has a horseshoe-like structure in the dust continuum emission.
In particular, \citet{cas15} have studied multifrequency observations of HD 142527, from 34 to 700 GHz and revealed that the optical depth distribution also shows the horseshoe-like structure as similar to the continuum intensity distribution. The peak region at wavelengths shorter than $\sim1$ mm are optically thick ($\tau_{345 GHz}\gtrsim2$).
Such lopsided dust disks are interpreted as grain accumulations by a dust trap in the continuum emission regions, which are formed by a gas large-scale vortex \citep{zhu14} that captures large dust particles (about a millimeter in size) due to azimuthal density gradients in the gas.

\section{Observations}

The 0.87 mm ($=870$ $\mu$m) ALMA dust polarization observations were carried out on 2016 March 11 during its Cycle 3 operation and on 2017 May 21 during its Cycle 4 operation.
The antenna configurations were  C36-2/3 with 38 antennas and C40-5 with 45 antennas, respectively.
A total of four spectral windows (spws) were set in the lower (2 spws) and upper (2 spws) sidebands, with 64 channels per spw and a 31.25 MHz channel width, providing a bandwidth of $\sim7.5$ GHz in total.
The central frequencies in the spws were 336.5, 338.5, 348.5, and 350.5 GHz, respectively.
The bandpass and gain calibrations were performed using observations of J1427-4206 and J1604-4441,
respectively, and the polarization calibration was performed using observations of J1512-0905 and J1427-4206.
The polarization calibrator was observed $3-4$ times with $\sim6$ min integration time during each observation schedule for calibration of the instrumental polarization ({\it D}-terms), cross-hand delay, and cross-hand phase.
The total integration times for the target were 73 min in Cycle 3 operation and 80 min in Cycle 4 operation.
The data taken on 2016 March 11 were previously reported by \citet{kat16}. The beam size was $0\farcs51\times0\farcs44$.
The new delivered data taken on 2017 May 21 are reported in this paper for the first time. The resolution of these data is higher than that of previous observations.
The reduction and calibration of the data were done with CASA version 4.5.3 \citep{mcm07} in a standard manner.
A detailed description of the data reduction is given by \citet{nag16}.

All images were reconstructed using the CASA task tCLEAN, with Briggs weighting with a robust parameter of 0.5 by combining these data.
The previous and new data were combined in the uv plane in the tCLEAN process.
In addition, to improve the sensitivity and image fidelity, self-calibration for both phase and amplitude was performed.
The beam size of the final product was $0\farcs27\times0\farcs24$, corresponding to a spatial resolution of $\sim38\times34$ au at the assumed distance of 156 pc.
An image with the highest spatial resolution using a robust parameter of $-2$ is shown in Appendix \ref{sec:Apn1}.

Stokes {\it Q} and {\it U} components give the total polarized intensity $PI=\sqrt{Q^2+U^2}$. Note that we ignore the Stokes {\it V} component in this study because it has not been well characterized for ALMA. 
The $PI$ value has a positive bias because it is always a positive
quantity. This bias has a particularly significant effect in low-signal-to-noise measurements. We thus debiased the polarized intensity map as
described by \citet{vai06} and \citet{hul15}.
The rms noise of the polarized intensity is derived to be $\sigma_{\rm PI}=2.9\times10^{-5}$ Jy beam$^{-1}$. The 1$\sigma$ error of the polarization angle is derived to be $\sim2.5^\circ$ on average.
The error of the polarization angle is taken from the CASA guide\footnote{3C~286 polarization CASA guide:\url{https://casaguides.nrao.edu/index.php/3C286\_Polarization}}  and is calculated as
\begin{equation}
\sigma_{\rm PA}({\rm ^\circ}) = 0.5\times 180/\pi\times \Big(\sqrt{(U\times\Delta Q)^2+(Q\times\Delta U)^2}/PI^2\Big).
\end{equation}

The polarization fraction ($P_{\rm frac}=PI/I$) is derived only where the detection is above the threshold 3$\sigma_{\rm PI}$.

\section{Results}

\subsection{New High Spatial Resolution Data}

\begin{figure}[htbp]
  \begin{center}
  \includegraphics[width=8.5cm,bb=0 0 2660 2215]{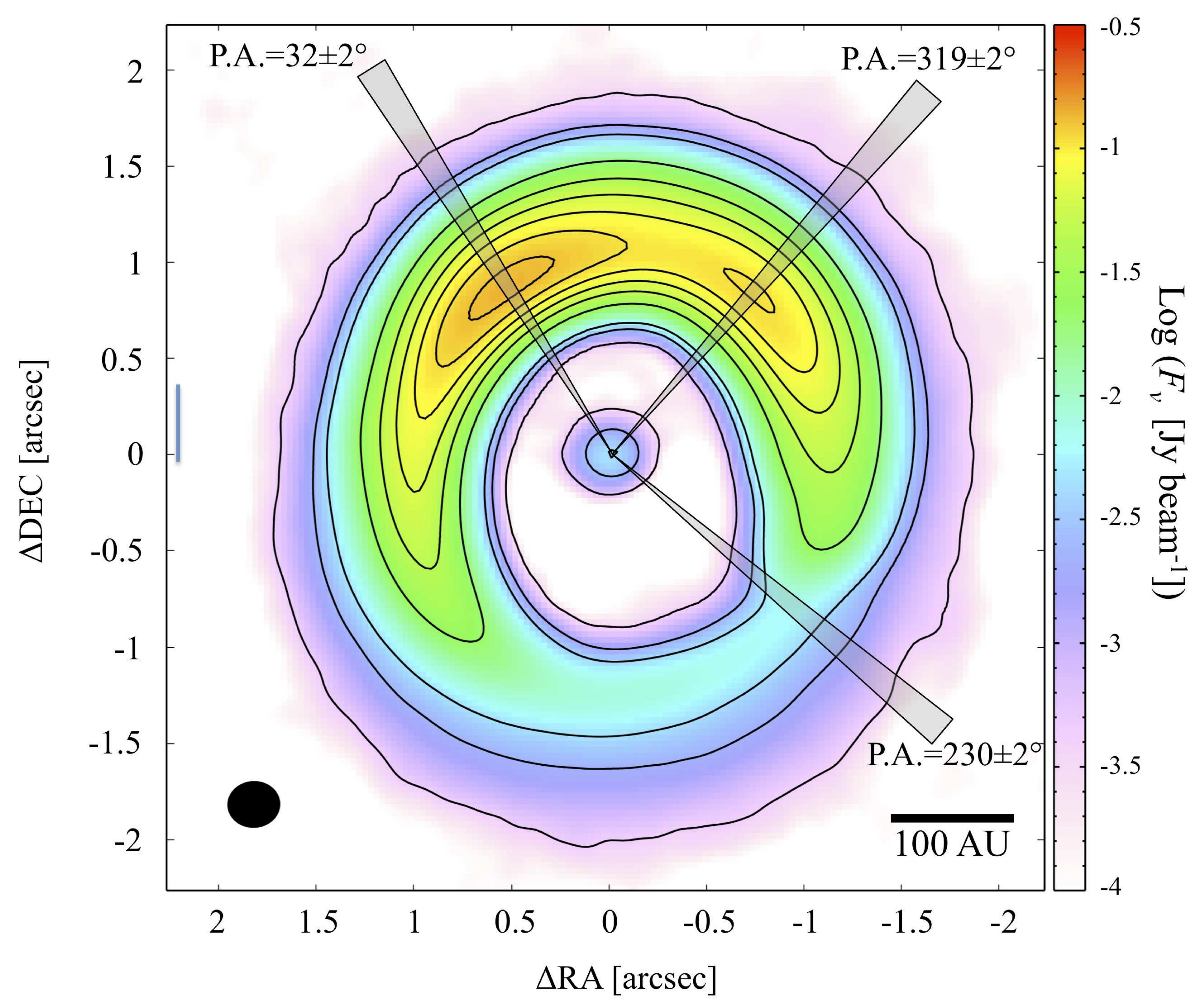}
  \end{center}
  \caption{Total intensity (Stokes {\it I}) map of the continuum emission at 0.87 mm with Briggs weighting with a robust parameter of 0.5. The synthesized beam with a size of $0\farcs27\times0\farcs24$ and a position angle of P.A. $=-86.6^\circ$, is indicated by the filled ellipse in the bottom-left corner.
  The contours correspond to $(10,50,100,500,1000,1500,2000,2500,3000)\times\sigma_I$, where $\sigma_I$ is $4.3\times10^{-5}$ Jy beam$^{-1}$.
  The areas where the azimuthal average is taken to obtain the profiles are indicated by gray shadows. 
  }
  \label{fig1}
\end{figure}

Figure \ref{fig1} shows the total intensity (Stokes {\it I}) map of the circumstellar disk around HD 142527, showing the horseshoe-like structure, which has been reported in the previous studies \citep{cas13,fuk13,cas15,boe17}.
There are two local peaks at P.A. $=32^\circ$ and $319^\circ$ in the northern region, and the minimum intensity on the ridge is found at P.A. $=230^\circ$ in the southwestern region.
The position angle is measured counterclockwise (east) from North.
The directions of the local peaks and the minimum intensity are shown in Figure \ref{fig1}.
The total flux is derived to be 3.0 Jy, which is consistent with the previous study \citep[][]{boe17}.

\citet{boe17} reported a compact source inside the dust cavity at $\sim50$ au North from the primary star.
However, we find no such source in the cavity.
The high resolution image shown in Appendix \ref{sec:Apn1} also shows no source in the cavity.
The compact source inside the cavity may be noise emission as our observations have better spatial resolution (a beam size of $0\farcs27\times0\farcs24$) and better sensitivity (a rms of $\sim43$ $\mu$Jy beam$^{-1}$) compared with those of \citet{boe17}, where the beam size was $0\farcs27\times0\farcs31$ and the rms value was $\sim115$ $\mu$Jy beam$^{-1}$. 

\begin{figure*}[htbp]
  \begin{center}
  \includegraphics[width=18cm,bb=0 0 2879 1202]{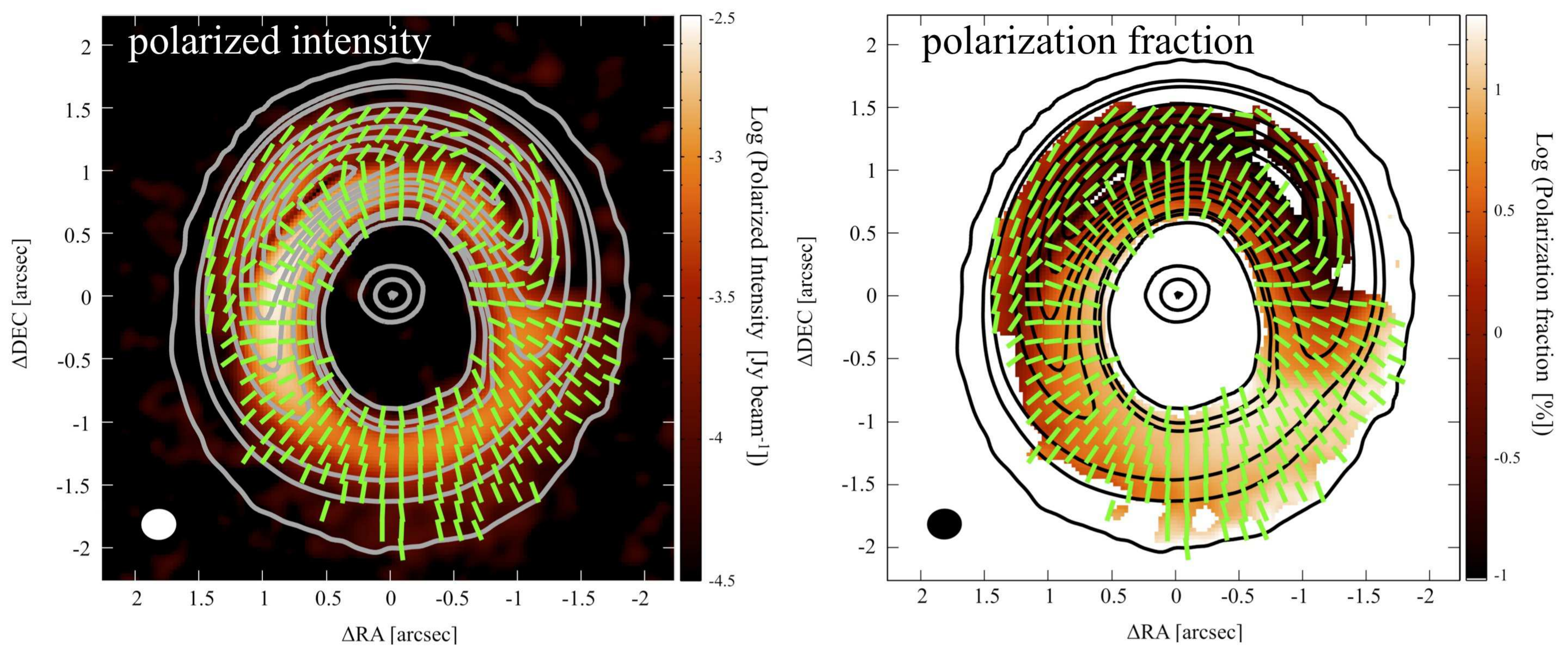}
  \end{center}
  \caption{Polarized intensity and polarization fraction maps color-corded on a logarithmic scale. The contours show the continuum emission, and the green vectors are the polarization vectors. The polarization vectors and the polarization fraction are shown where the polarized intensity is higher than 3$\sigma_{\rm PI}$. 
Note that the lengths of the polarization vectors are set to be the same.
  }
  \label{fig9}
\end{figure*}

\begin{figure}[htbp]
  \begin{center}
  \includegraphics[width=8.5cm,bb=0 0 2462 2248]{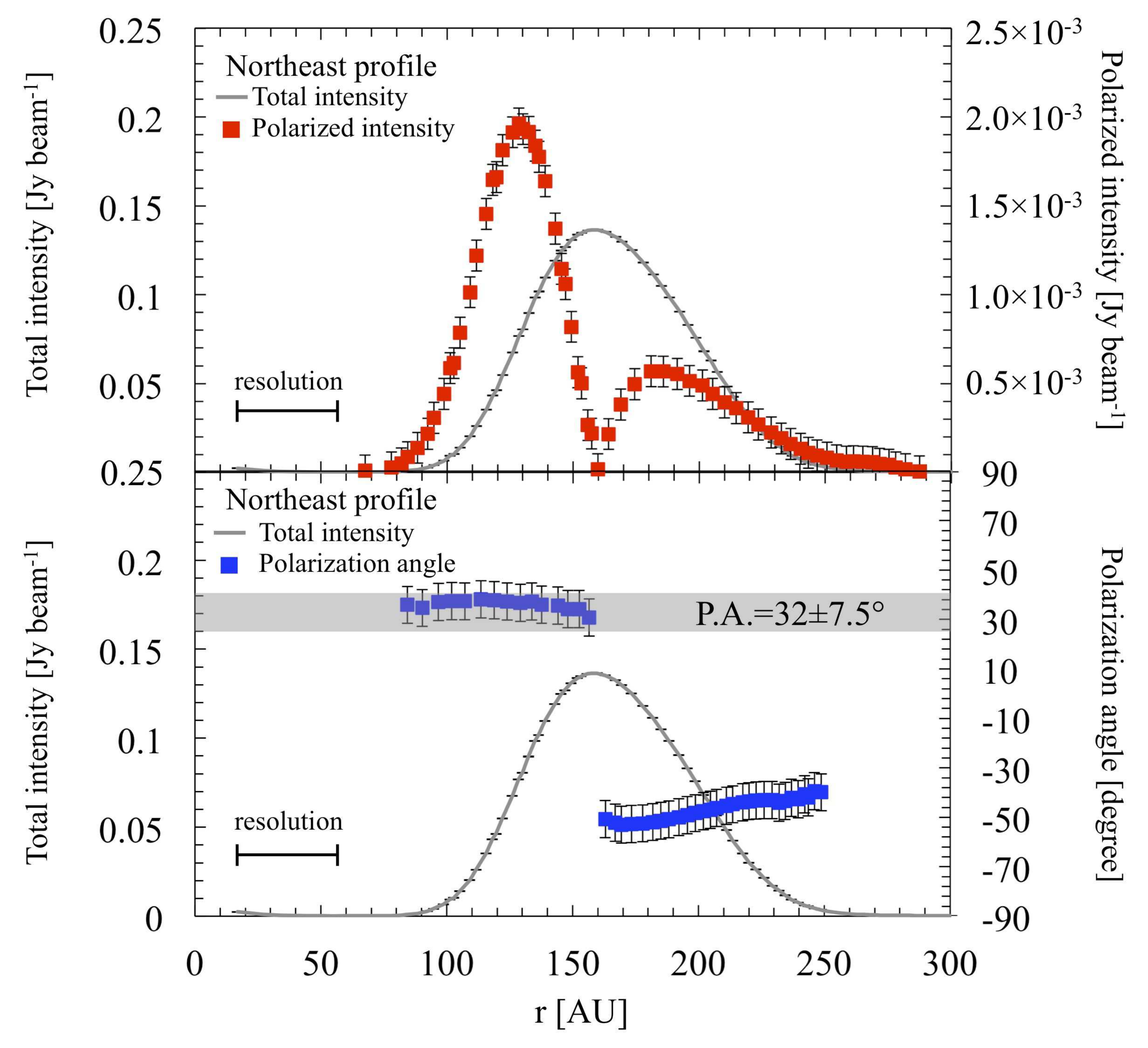}
  \end{center}
  \caption{Upper panel: Radial profile of the total intensity (Stokes {\it I}, gray line) and the polarized intensity (red squares) at 0.87 mm for P.A. $=32\pm2^\circ$. Lower panel: Radial profile of the total intensity (Stokes {\it I}, gray line) and the polarization angle (blue squares) for P.A. $=32\pm2^\circ$. The gray shadow represents the position angle. 
  The horizontal bar denotes a spatial resolution of 40 au, corresponding to $\sim0\farcs26$. The error bar represents 3$\sigma$ ($3\sigma_{\rm I}=1.3\times10^{-4}$ Jy beam$^{-1}$, $3\sigma_{\rm PI}=8.7\times10^{-5}$ Jy beam$^{-1}$, and  $3\sigma_{\rm PA}=7.5^\circ$).
  }
  \label{plot1}
\end{figure}

\begin{figure}[htbp]
  \begin{center}
    \includegraphics[width=8.5cm,bb=0 0 2462 2248]{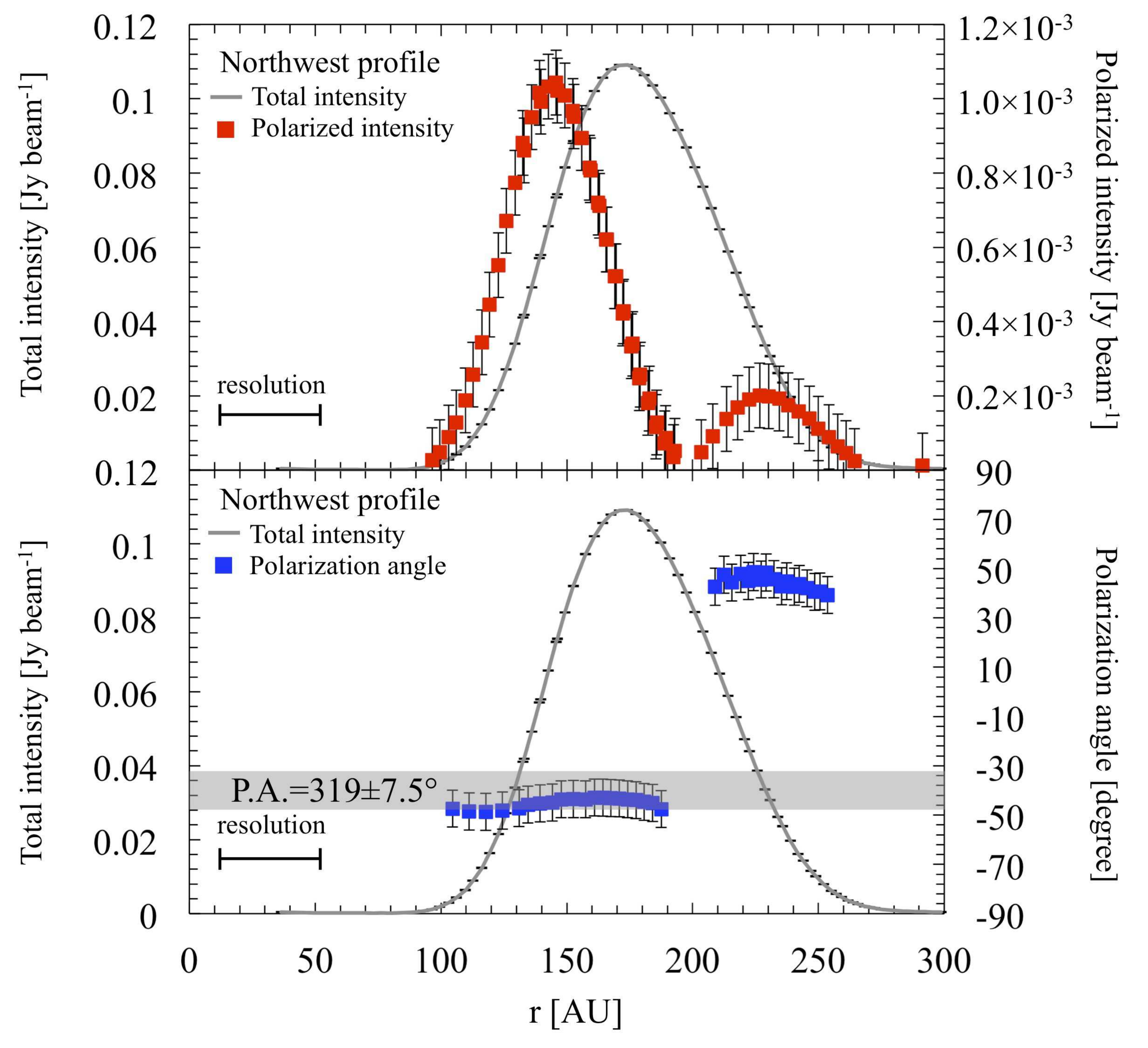}
  \end{center}
  \caption{Same as Figure \ref{plot1} but for P.A. $=319\pm2^\circ$
  }
  \label{plot2}
\end{figure}

\begin{figure}[htbp]
  \begin{center}
  \includegraphics[width=8.5cm,bb=0 0 2462 2248]{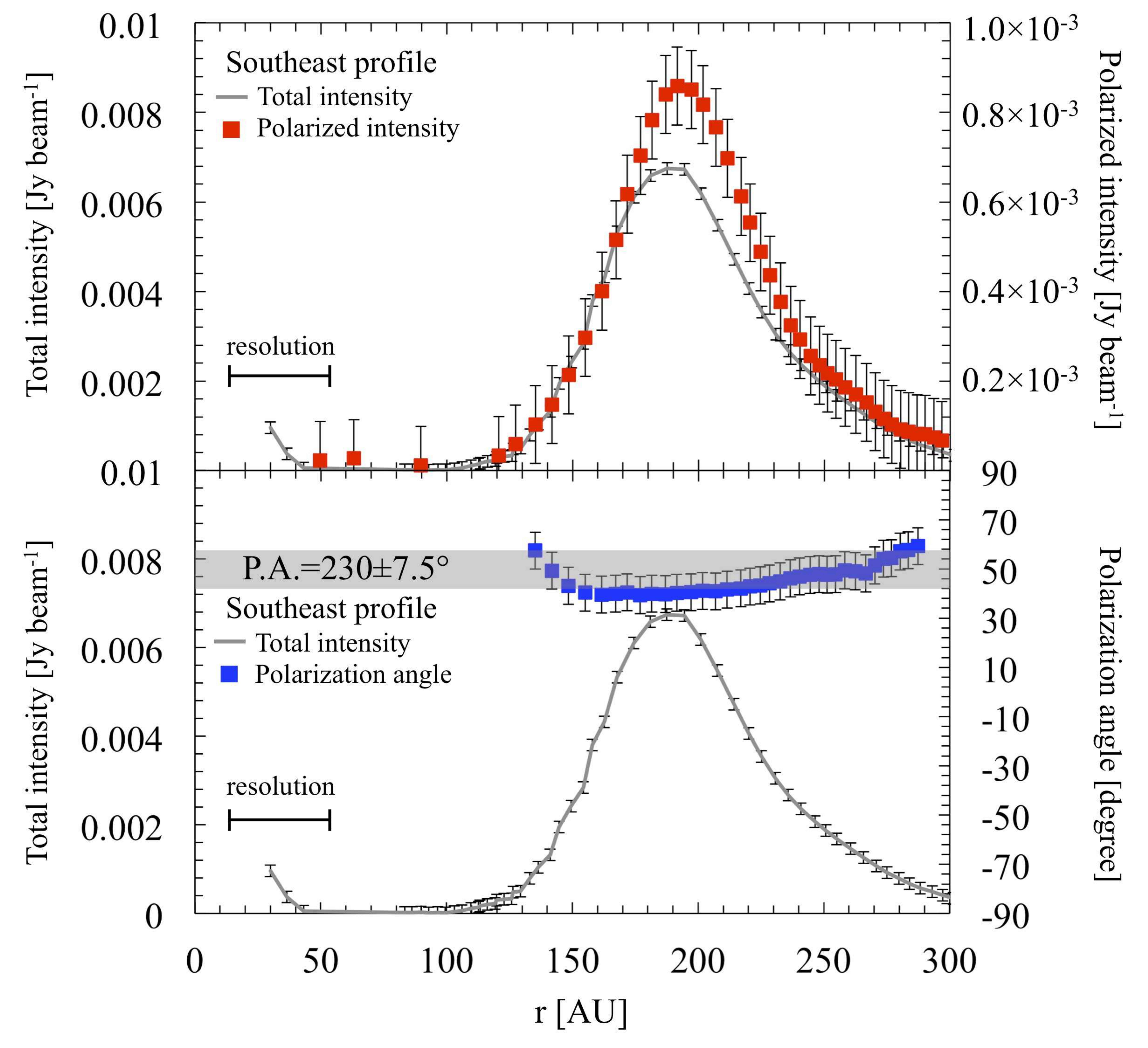}
  \end{center}
  \caption{Same as Figure \ref{plot1} but for P.A. $=230\pm2^\circ$
  }
  \label{plot3}
\end{figure}

Figure \ref{fig9} shows the polarized intensity and the polarization fraction overlaid with the total intensity and the polarization vectors.

The polarized intensity shows not only a ring-like distribution but also an additional component outside the ring in the northern region.
The polarization vectors on the ring point in radial directions, whereas those outside the northern ring point in azimuthal directions.
These structures and the flip of the polarization vectors are previously reported by \citet{kat16}.
We resolve the ring structure of the polarized intensity and clearly observe the flip of the polarization vectors because the spatial resolution of the new data is twice that of previous data.
The flip is found from P.A. $\sim250^\circ$ to $90^\circ$ (in a counterclockwise direction).
We note that the flip of the polarization vectors appears only in the northern region of the horseshoe structure.
The southern region only shows polarization vectors pointing in radial directions.
We also find that the polarized intensity distribution in the southern region is proportional to the total intensity distribution. This was not found by \citet{kat16} due to the lower spatial resolution of their data.

To compare the total intensity and polarized intensity profiles, and show the flip of the polarization angles clearly, we plot the total intensity, polarized intensity, and polarization angle as a function of the radial cut along the position angles of P.A. $=32\pm2^\circ,230\pm2^\circ$, and $319\pm2^\circ$ in Figures \ref{plot1} $-$ \ref{plot3}, respectively.
These position angles correspond to the local peak and minimum intensity directions shown in Figure \ref{fig1}.

Along the radial cuts at P.A. $=32\pm2^\circ$ and $319\pm2^\circ$ (Figures \ref{plot1} and \ref{plot2}), the total intensity distribution has a single peak and the polarized intensity distribution shows a double-peak profile. 
The inner peaks of the polarized intensity are four times brighter than the outer peaks.
The local minima of the polarized intensity coincide with the peaks in the total intensity where high optical depths are obtained from the observations of ALMA band 7 ($\sim870$ $\mu$m) and band 9 ($\sim430$ $\mu$m)  \citep[$\tau\gtrsim2$ at 345 GHz;][]{cas15}.
The physical reasons for the steep drops in the polarized intensity are discussed in the next subsection.

Figures \ref{plot1} and \ref{plot2} also show that the polarization vectors are rotated by $90^\circ$ at or just outside the total intensity peaks.
The $90^\circ$ flip of the polarization angle occurs sharply, and no polarization vectors point in intermediate directions.
In the northeast profile (Figure \ref{plot1}), we cannot resolve the boundary region where the flip occurs, indicating that the flip of the polarization vectors occurs at a scale that is smaller than the beam size.
In the northwest profile (Figure \ref{plot2}), we find that the flip of the polarization vectors has a gap of $\sim20$ au, where no polarization is detected.

Along the radial cut that includes the minimum intensity on the ridge toward P.A. $=230\pm2^\circ$ (Figure \ref{plot3}), the total intensity and polarized intensity distributions show quite similar (Gaussian-like) profiles.
The polarization angle is constant, around $\sim50^\circ$ in the region $r=120-300$ au. 
The 90$^\circ$ flip of the polarization vector is not found in this radial cut.

Figure \ref{fig9} also shows the polarization fraction distribution.
We are unable to determine the polarization fraction at the local continuum peaks because no polarization is detected at these local peaks.
A high polarization fraction of $\sim15$\% is found in the minimum intensity region (P.A. $\sim230^\circ$).
This high polarization fraction in the southern region was previously reported by \citet{kat16}. Our observations support these results at higher spatial resolution.

At the position of the star, the continuum emission from the inner disk is detected with a peak intensity of 4.4 mJy beam$^{-1}$ even though it is still spatially unresolved. However, no polarization is detected in this region. We calculate the upper limit of the polarization fraction to be $\sim2.0$\% by assuming a polarized intensity of 87 $\mu$Jy beam$^{-1}$, corresponding to the 3$\sigma$ noise level of the polarized emission.

\subsection{Dependence on Spatial Resolution}
\begin{figure*}[htbp]
  \begin{center}
  \includegraphics[width=17cm,bb=0 0 2580 2248]{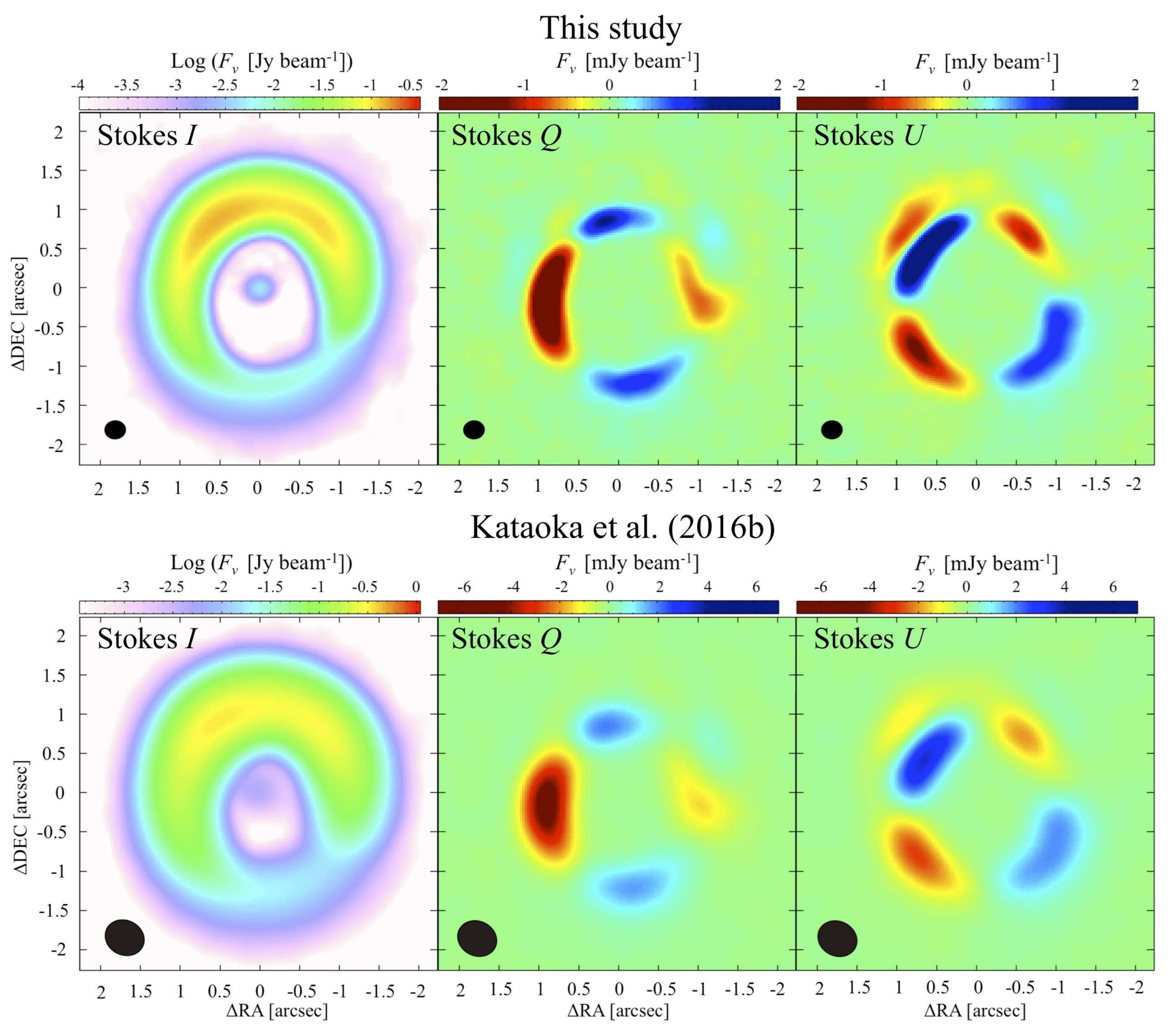}
  \end{center}
  \caption{ALMA observations of the Stokes parameters of {\it I}, {\it Q}, and {\it U}. The upper and lower panels show images based on our data and data used by \citet{kat16}, respectively. The color magnitudes are scaled to indicate the same intensity in both data sets.
  }
  \label{fig:comp}
\end{figure*}

We compare our high spatial resolution data with data previously reported by \citet{kat16} to investigate the consistency and dependence on spatial resolution.
Figure \ref{fig:comp} shows the Stokes {\it I}, {\it Q}, and {\it U} images based on our data and previous data, respectively. The color magnitudes are scaled to indicate the same intensity in both data sets.
The Stokes {\it I}, {\it Q}, and {\it U} images based on the two observations look similar.
However, the disk structure is resolved more clearly with the new data thanks to their better spatial resolution.
For example, we identify a cavity hole inside the outer disk in the Stokes {\it I} image; this cavity unclear and diluted by the larger beam in the image based on previous data.

The Stokes {\it Q} and {\it U} change their signs along the radial direction in the northern disk, which corresponds to the flip of the polarization vectors.
In the northeast direction of the Stokes {\it U} map, the location of the flip, i.e., where the Stokes {\it Q} and {\it U} vanish, cannot be resolved even with the new data, and the location of the boundary is slightly different from that obtained using previous data. The images based on new data show that the boundary is located $\sim10$ au inside closer to the center compared to that in the images based on previous data.
As a result, the flip of the polarization vectors occurs at smaller radii in our observations.
The difference in the location of the boundary between the previous and new data is caused by the beam smoothing effect since the {\it Q} and {\it U} values change from positive to negative with increasing radius. The shape and the boundary region are changed by the beam smoothing effect.
At P.A. $=32^\circ$ in the {\it U} maps, the positive and negative peaks of our data are almost two times higher than those of the previous data after the intensity is corrected for beam size. These results suggest that beam dilution occurred in previous observations and may also affect our observations.

In the southern region, the sign of the Stokes {\it Q} or {\it U} does not change along radial directions from the central position, and there is no flip of the polarization vector.
At P.A. $=230^\circ$ in the {\it U} maps, the positive peak of our data is 1.5 times higher than that of the previous data after the intensity is corrected for beam size. The southern region may be less affected by beam dilution than the northern region.

Here, we compare the polarization fraction for our observations with that for the previous observations to investigate the depolarization effect and the dependence of the polarization fraction on the spatial resolution.
In general, larger beam observations yield a lower polarization fraction if the polarization direction varies within the beam (depolarization effect).

The polarized intensity and polarization fraction are useful to investigate the depolarization effect.  We derive the polarized intensity from the Stokes {\it Q} and {\it U} values. In order to make the polarized intensity for our observations the same as that for the previous observations, the Stokes {\it Q} and {\it U} images of our data need to be smoothed to match those of the previous data. Then, the polarized intensity should be derived from the smoothed Stokes {\it Q} and {\it U} data. If the polarized intensity of our data ($PI=\sqrt{Q^2+U^2}$) is smoothed to match that of the previous data, the smoothed polarized intensity of our data may give different values from that of the previous data because the polarized intensity is the combination of the Stokes {\it Q} and {\it U} values, and is always a positive quantity.
Only if the polarization pattern has a uniform distribution or is fully resolved by the previous observations (the depolarization does not happen), the smoothed polarized intensity is the same as that of the previous data.
Therefore, it is possible to determine whether the polarization pattern was fully resolved by the previous observation by deriving the ratio of the polarization fraction.

To assess the depolarization effect, the polarized intensity ($PI=\sqrt{Q^2+U^2}$) and the total intensity of our data are smoothed to match those of the previous data. The polarization fraction is derived by using the smoothed polarized and total intensity data. Then, we calculate the ratio of the polarization fraction for our high resolution data to that for the previous data.  
We confirm that the polarization fraction is consistent within $\sim30$\% between our data and previous data if the polarized intensity is derived from the Stokes {\it Q} and {\it U} data which are smoothed to those of the previous data.

\begin{figure}[htbp]
  \begin{center}
  \includegraphics[width=8.5cm,bb=0 0 2682 2220]{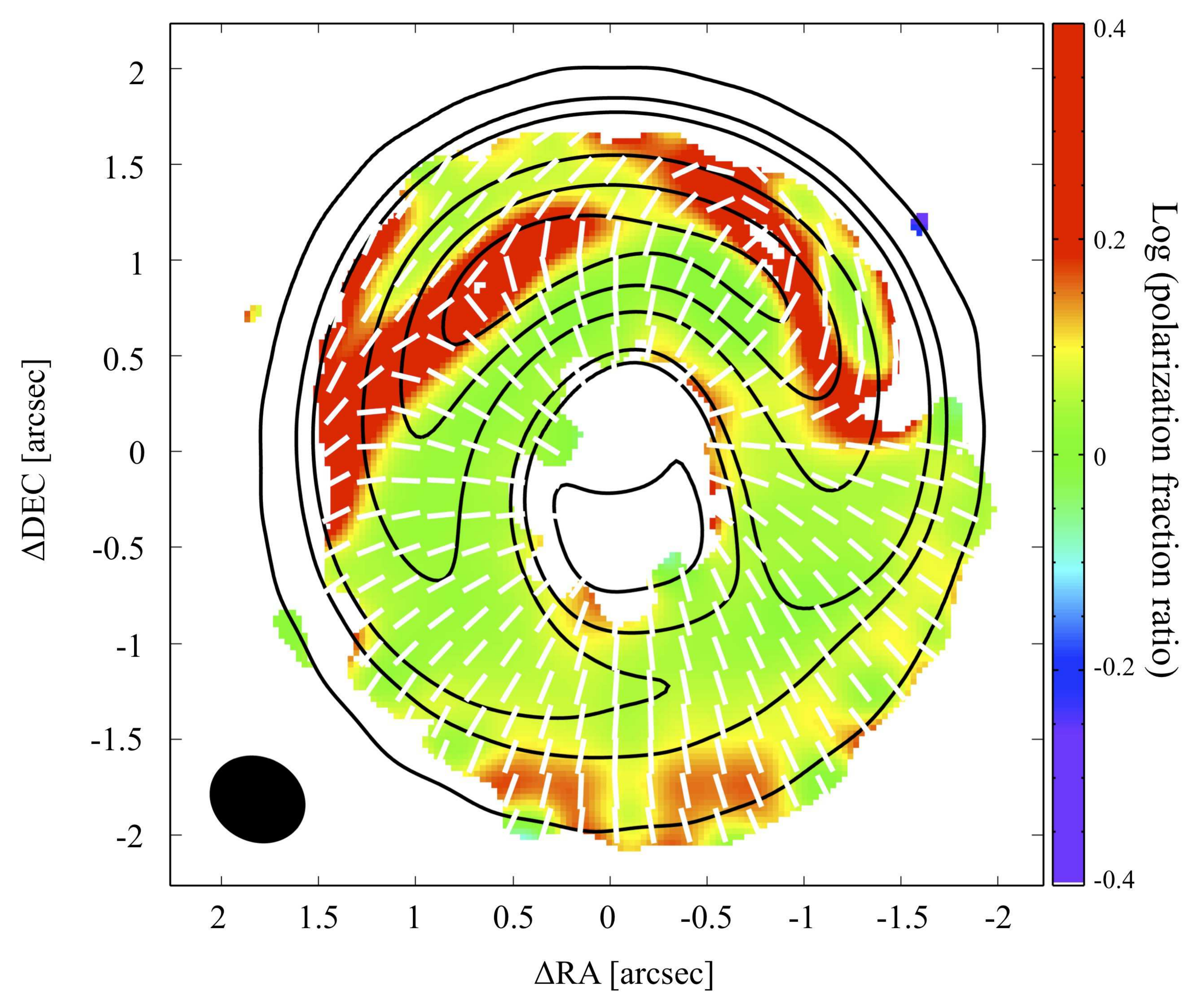}
  \end{center}
  \caption{Ratio of the polarization fraction for our observations to that for previous observations, done by \citet{kat16}, color-coded on a logarithmic scale.
  Our observations are analyzed with a broadened beam to make the angular resolution the same as that of \citet{kat16}.
  The beamsize ($0\farcs51\times0\farcs44$) is shown by the filled ellipse in the bottom left corner. The contours correspond to $(10,50,100,500,1000,1500)\times\sigma_I$, where $\sigma_I$ is $185$ $\mu$Jy beam$^{-1}$.
  }
  \label{fig6_1}
\end{figure}

Figure \ref{fig6_1} shows the ratio for the polarization fraction of our high resolution data (smoothed to the previous beam size) to that for the previous data.
In the northern region, where the flip is found, the ratio of the polarization fraction is $\gtrsim0.4$ on a logarithmic scale ($\gtrsim2.5$ on a linear scale), indicating that our high resolution observations resolve the change of the polarization angles and give the correct polarized intensity. In contrast, the previous observations exhibit a depolarization effect due to the larger beam size, leading to a reduced polarization fraction.
In the whole region excepting for the part where the flip is found, the ratio is almost unity, which indicates that the patterns of the polarization vectors are ordered and were resolved by the previous observations. Therefore, the polarization patterns are correctly observed without the depolarization effect in both our data and previous data excepting for the region where the flip is found. Observations with higher spatial resolution may be important for the part where the flip is found.

\begin{figure}[htbp]
  \begin{center}
  \includegraphics[width=8.5cm,bb=0 0 2682 2220]{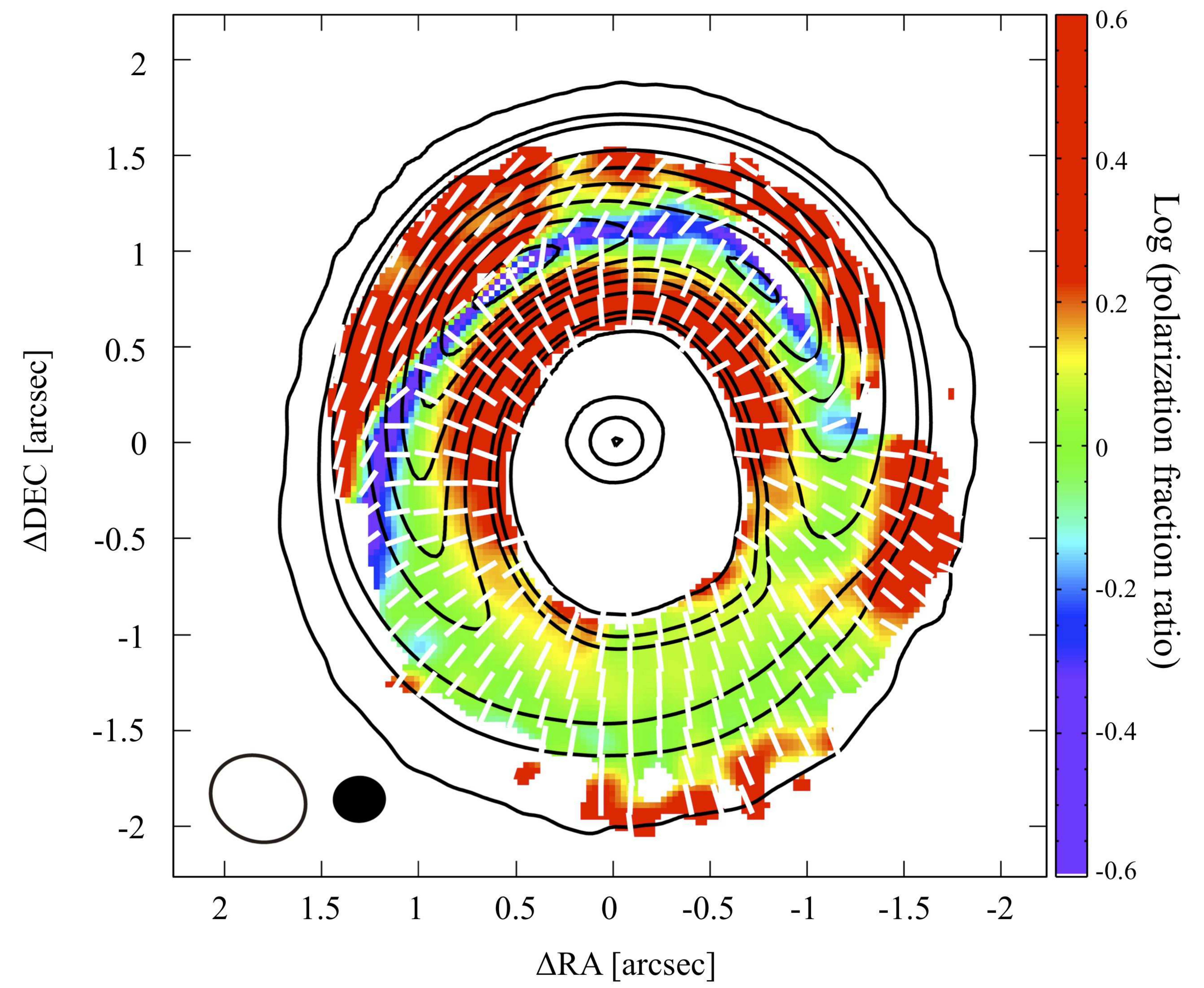}
  \end{center}
  \caption{Ratio of the polarization fraction for our observations to that for previous observations, done by \citet{kat16} color-coded on a logarithmic scale.
  The beam sizes in the previous observations ($0\farcs51\times0\farcs44$) and our observations ($0\farcs27\times0\farcs24$) are shown by the white ellipse and filled ellipse, respectively in the bottom left corner. The contours are the same as those in Figure \ref{fig1}.
  }
  \label{fig6}
\end{figure}

We also derive the ratio of the polarization fraction for our high resolution data without smoothing to that for the previous data.
Using this ratio, we investigate the dependence of the polarization fraction on the spatial resolution.
Figure \ref{fig6} shows the ratio of the polarization fraction for our high resolution data to that for the previous data.
In the northern region, the variation of the ratios is $\pm0.6$ on a logarithmic scale ($\gtrsim4$ on a linear scale).
In particular, the local peaks in the total intensity map have large variations because the polarized intensity drops at these peaks.
In addition, the inner edge of the northern region of the disk shows a high ratio of the polarization fractions.
These results indicate that the polarization fractions are not constant over the beam size range of $\sim0\farcs2-0\farcs5$.

The variations of the ratio at the continuum peak can be explained by the depolarization effect and the high optical depth.
The high ratio of the polarization fractions is explained by the depolarization effect, as shown in Figure \ref{fig6_1}. 
The low ratio of the polarization fractions at the peak shown in Figure \ref{fig6} may be due to the high optical depth.
Polarization from self-scattering and that from radiative grain alignment are both caused by anisotropic radiation.
Therefore, the polarization fraction in opaque regions decreases with optical depth because the radiation field becomes isotropic due to multiple scattering and absorption.
The polarization of the grain alignment with magnetic fields also decreases with optical depth because the radiation gradients are important for grain alignment.
The previous observations did not resolve the high optical depth regions and the polarization fraction was smoothed.
In contrast, our high resolution observations identify the region where the polarized intensity drops.
Therefore, we find the low ratio in these regions because of the high optical depth and the high spatial resolution.

\citet{yan17} discussed the effect of optical depth for both self-scattering and grain alignment assuming an inclination angle of $45^\circ$. They showed that with self-scattering, polarization decreases with optical depth once $\tau\gtrsim1$ (it increases until $\tau\sim1$) but it exists even at a high optical depth ($\tau\gtrsim4$).
In contrast, aligned grains cannot produce polarized emission in a high optical depth region.
According to these optical depth dependences, the drops of the polarized intensity might prefer the grain alignment. However, they assumed an inclination angle of $45^\circ$.  An inclined disk may have caused polarization due to self-scattering even at a high optical depths. In contrast, a face-on disk, such as our target, may produce an isotropic radiation field and the polarization may decrease due to self-scattering.
Actually, scattering is suggested to become less dominant with decreasing an inclination \citep[][]{yan16b,yan17}.
Another possibility is that the depolarization occurs at a smaller scale than our high spatial resolution. The polarization may be canceled out by mixing the two orthogonal polarization vectors within the beam size. Therefore, the polarized intensity drops at the position where the flip occurs.

The inner edge of the disk in Figure \ref{fig6} also shows a high ratio, indicating that the polarization fraction increases with increasing spatial resolution. This may mean that the polarized intensity has a steeper gradient than that of the total intensity, assuming that the total intensity is optically thin at the inner edge.
We discuss the dependence of the polarization fraction on spatial resolution in Section 4.2.

\section{Discussion}
We discuss the possible mechanisms of the polarization of the protoplanetary disk of HD 142527 in this section.
Three mechanisms for polarization in protoplanetary disks have been proposed: (1) grain alignment with magnetic fields \citep[e.g.,][]{cho07}, (2) grain alignment with radiation gradients \citep[e.g.,][]{taz17}, and (3) self-scattering of thermal dust emission \citep[e.g.,][]{kat15}.
These three mechanisms cause different patterns of polarization vectors and polarization degrees, as discussed in the following subsections.
We compare the observed polarizations with the polarization predicted by each theory and discuss the polarization mechanisms.

\subsection{Morphology of Polarization Vectors}

\begin{figure}[htbp]
  \begin{center}
  \includegraphics[width=8.5cm,bb=0 0 2998 1653]{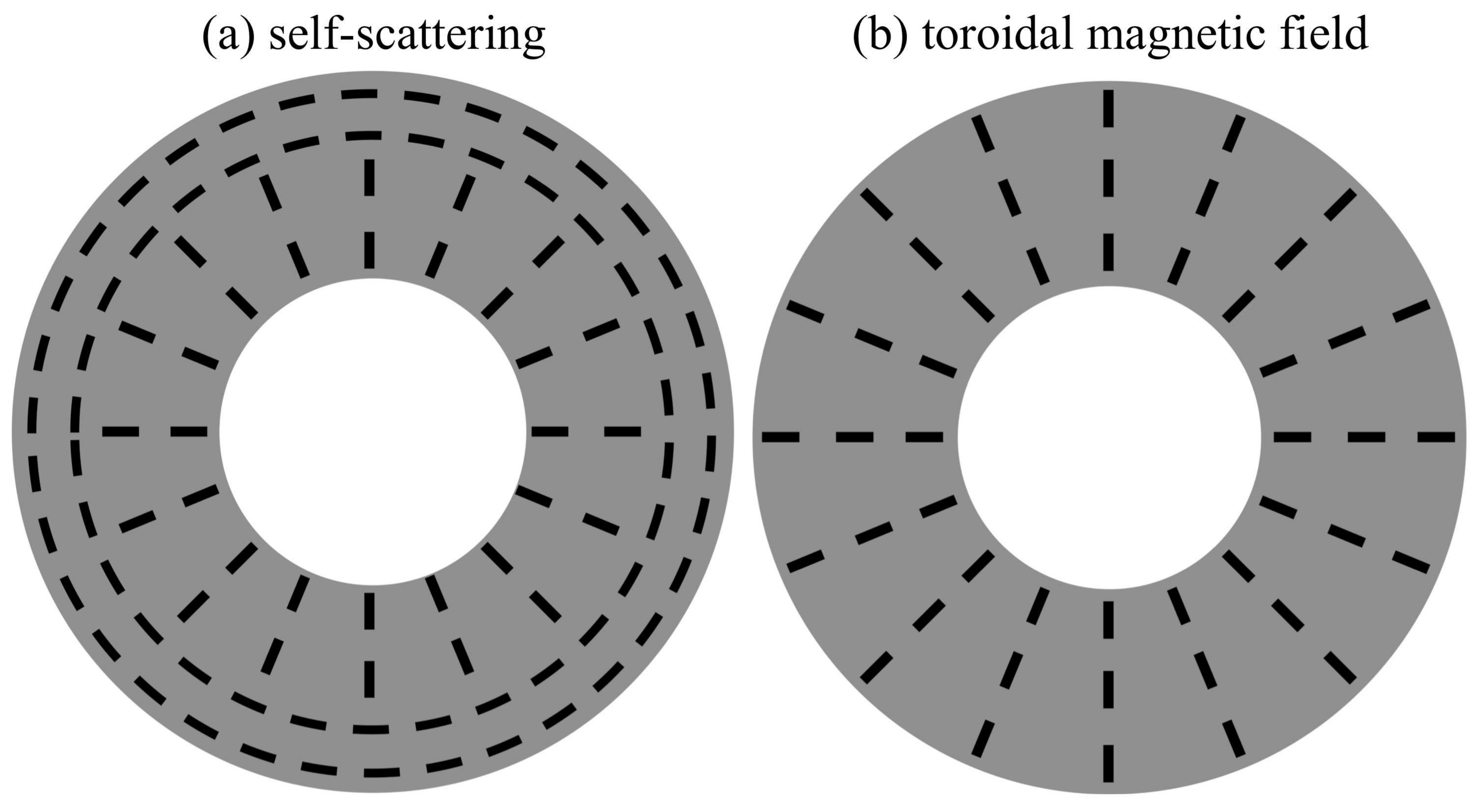}
    \end{center}
  \caption{Schematic views of polarization vectors predicted by two mechanisms of polarization of thermal dust emission in HD 142527.
  (a) Grain alignment with toroidal magnetic fields and (b) self-scattering of thermal dust emission.
  }
  \label{polari_view}
\end{figure}

We discuss whether the polarization in the disk is due to grain alignment or self-scattering by investigating the distributions of the polarization vectors.
Magnetic fields or radiation gradients can determine the direction of grain alignment.
Figure \ref{polari_view} shows schematic views of the polarization vectors predicted by self-scattering and grain alignment with magnetic fields in the disk of HD 142527. The theory of radiative grain alignment is discussed later.

\begin{figure*}[htbp]
  \begin{center}
  \includegraphics[width=17cm,bb=0 0 2998 1931]{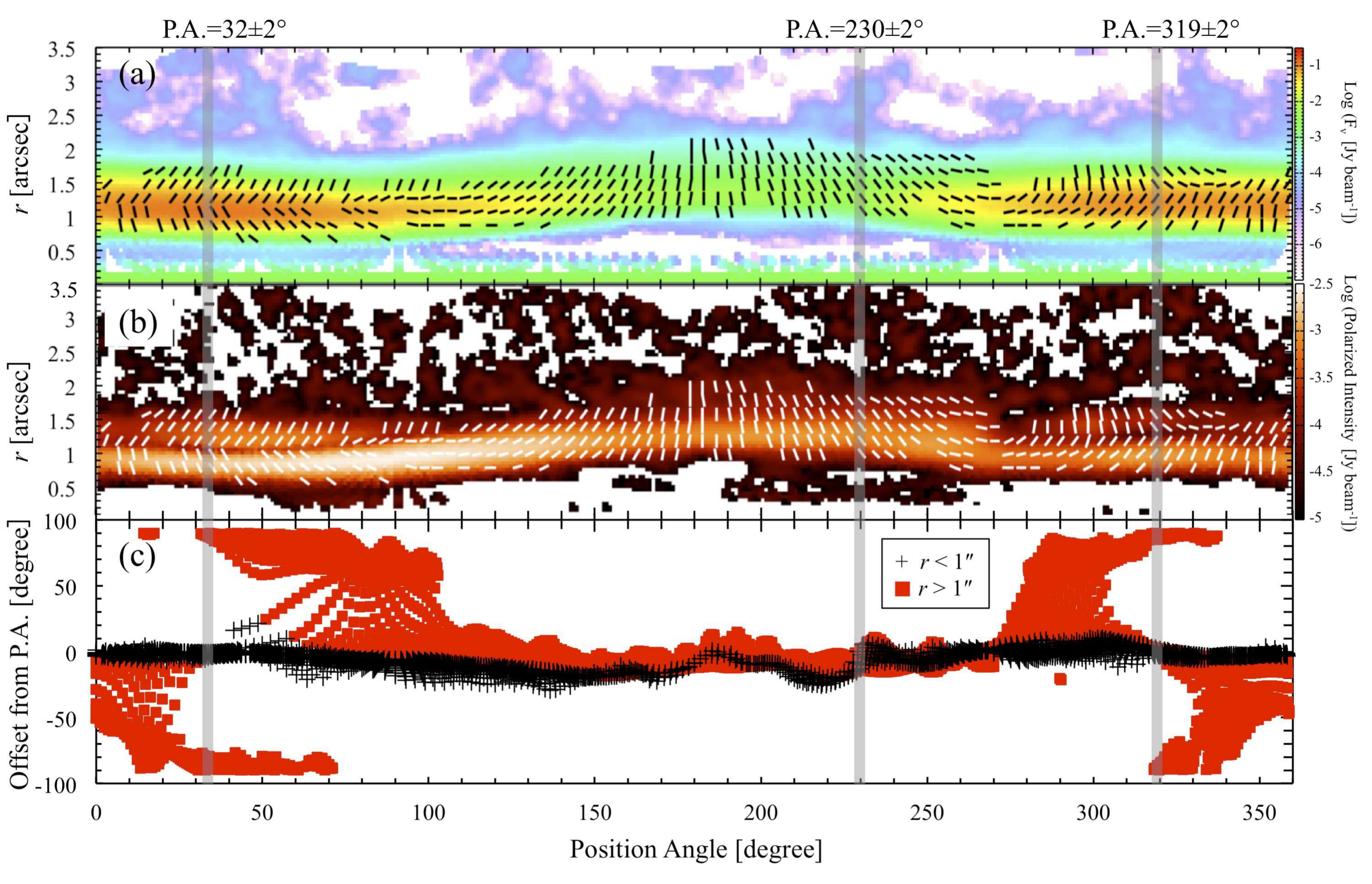}
    \end{center}
  \caption{Upper panel (a) shows the total intensity as a function of the radius from the central star and of the position angle overlaid with the polarization angles. Middle panel (b) shows the same but for the polarized intensity. 
Bottom panel (c) shows the offset between the observed polarization vectors and the radial direction of the disk as a function of position angle.
  The black crosses represent the offset within the radius of $r<1\arcsec$, and the red squares represent that outside the radius of $r>1\arcsec$. The gray shadows represent the position angles, as in Figure \ref{fig1}.
  }
  \label{plot8}
\end{figure*}

For self-scattering, \citet{kat15,kat16} derived the expected distributions of the polarization vectors in the disk of HD 142527. The flip of the polarization vectors should occur outside the entire disk because the anisotropy of the radiation field changes with radius from the central star (Figure \ref{polari_view}$-$a).
The flip of the polarization vectors in the northern region can be explained by thermal dust scattering.
As shown by Figure 3 of \citet{kat16}, self-scattering predicts that a flip should also occur in the southern region, similar to that in the northern region, because the northern and southern intensity profiles are similar (both Gaussian-like). 
\citet{kat16} show that the flip should be detected at $r\sim1\farcs2-1\farcs5$ due to the changing of the anisotropic radiation field. The anisotropy of the radiation in the disk is calculated by a previous modeling of the continuum emission at the same wavelength \citep{mut15}.
However, no flip is observed in the southern region although a signal-to-noise ratio of the polarized intensity is low.
Therefore, it may not be likely that self-scattering is the mechanism responsible for polarization in the southern region.

Another polarization mechanism is grain alignment with magnetic fields. 
The magnetic fields in protoplanetary disks have been studied using magnetohydrodynamics (MHD) simulations; a toroidal distribution due to magnetorotational instability and differential rotation effects has been reported \citep[e.g.,][]{bra95,fro06,hen09,dav10,bai13,suz14}.
Therefore, toroidal magnetic fields in disks are usually assumed to cause aligned polarization vectors due to magnetically aligned grains \citep[e.g.,][]{cho07,ber17}.
If a disk has toroidal magnetic fields, the polarization vectors will point in radial directions (Figure \ref{polari_view}$-$b).

Figures \ref{plot8} (a) and (b) show the total intensity and polarized intensity, respectively, as functions of the radius and of the position angle overlaid with the polarization orientations.
In these figures, the flip of the polarization vectors occurs in the ranges of P.A. $=0^\circ-100^\circ$ and P.A. $=270^\circ-360^\circ$.
The polarized intensity ring can be identified as the bright belt in Figure \ref{plot8} (b).
The polarization vectors on this ring rotate with increasing position angle.

To verify the locations of the flip of the polarization vectors, we calculate the offset angle between the polarization angle and the position angle measured from the central star.
We calculate the projected radial directions for each point, where the polarization vectors are plotted assuming an inclination of $27^\circ$ and a position angle (the major axis of the disk) of $341^\circ$  \citep{fuk13}.

We plot the offset of the polarization angle from the radial direction as a function of position angle in Figure \ref{plot8} (c).
The black dots represent the offsets at pixels where the distance from the central star is less than $1\farcs0$. 
The black dots are located in the offset angle between $-30^\circ$ and $20^\circ$ regardless of the position angle, which means that the polarization vectors in the inner region point in radial directions.
The red squares represent the offsets at a distance more than $1\farcs0$ from the central star. 
In the range of P.A. $=110-270^\circ$, the red squares are located in the offset angle between $-30^\circ$ and $20^\circ$, indicating that the polarization angles are consistent with the radial directions. However, in the ranges of P.A. $=0-90^\circ$ and P.A. $=270-360^\circ$, the red squares are located between $-90^\circ$ and $+90^\circ$, indicating a flip of the polarization vectors.

Taking into account the fact that the polarization angles are consistent with the radial directions in the range of P.A. $=110-270^\circ$, toroidal magnetic fields can explain the polarization vectors in these regions.
However, it is unlikely that the magnetic fields rotate by $90^\circ$ locally in the range of P.A. $=0-90^\circ$ and P.A. $=270-360^\circ$.
The change of the polarization vectors for aligned grains may be possible in the highly optically thick case due to absorption by aligned grains.
In this case, the polarization vectors may become parallel to the magnetic field direction \citep{yan17,har18}. However, we find it unlikely that the polarization flip is occurring due to optical depth for HD 142527. For example, polarization vectors at the inner edge of the northern region of the disk point in radial directions, whereas those at the outer edge point in azimuthal directions even though both sides are optically thin. 
Therefore, the flip of the polarization vectors is not due to the high optical depth in this disk. 
\citet{lee18} also find a change of the polarization vectors toward the young nearly edge-on protostellar disk in HH 111 (early Class I) protostellar system. They suggest that the polarization vectors change by reflecting the different regions of the nearside and farside of the disk. The polarization of the nearside may indicate toroidal magnetic fields, and that of the farside may indicate poloidal magnetic fields. 
In our case, the disk is almost face-on and the flip of the polarization vectors is found in the northern region.
Therefore, magnetic fields are not likely to explain the flip of the polarization vectors in our case.

For radiative grain alignment, no polarization modeling has been performed for the lopsided disk of HD 142527. 
Therefore, we construct a simple model to discuss the polarization vectors due to grain alignment with radiation gradients.
We first briefly summarize the mechanism of grain alignment with radiation gradients proposed by \citet{taz17}.
They calculated the alignment and precession timescales of dust grains and found that large dust grains ($\gtrsim$ a few 10 $\mu$m) are not aligned with the magnetic fields but with the radiation directions because the radiative precession time scale is dominant for grain alignment, and the Larmor precession timescale for grain alignment with magnetic fields becomes longer than the gaseous damping timescale. They found that the alignment axis is determined by the grain precession with respect to the radiative flux, indicating that the grain short axis is parallel to the radiation direction.
The polarization vectors of the dust thermal emission are parallel to the grain long axis and perpendicular to the radiation gradient.

\begin{figure}[htbp]
  \begin{center}
  \includegraphics[width=8cm,bb=0 0 2998 1567]{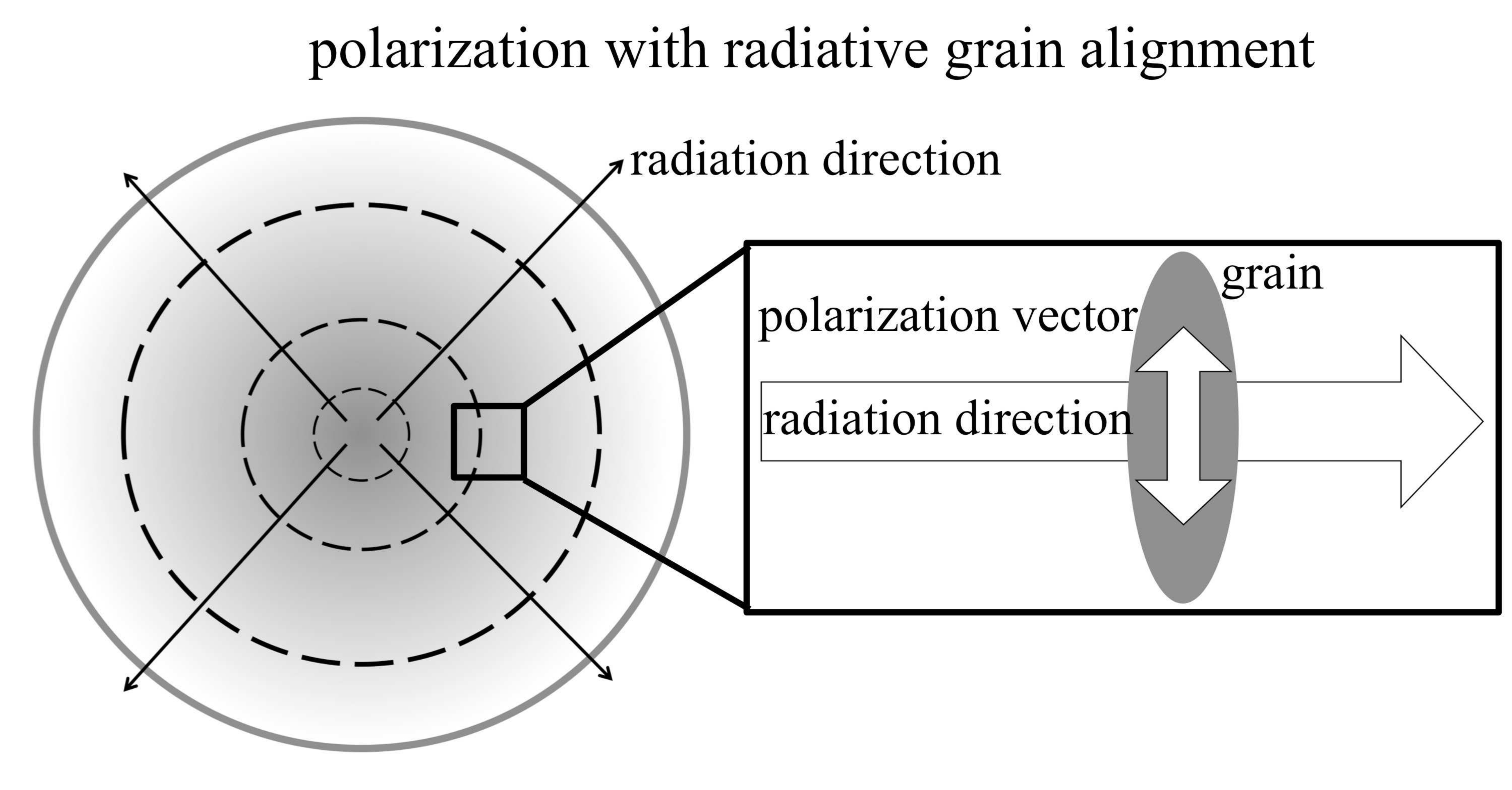}
    \end{center}
  \caption{Schematic view of polarization vectors for radiative grain alignment.
  }
  \label{rat_view}
\end{figure}

Before considering a lopsided disk, we consider a disk with an axisymmetric intensity distribution.
Figure \ref{rat_view} shows a schematic view of the polarization vectors with radiative grain alignment.
The intensity is assumed to increase toward the center.
The flux gradients point in the radial directions, and thus the polarization vectors point in azimuthal directions (i.e., the polarization vectors depend on the flux gradient).
Therefore, we can estimate the polarization vectors for radiative grain alignment by determining the radiation direction at each position.

\begin{figure*}[htbp]
  \begin{center}
  \includegraphics[width=18cm,bb=0 0 2677 1197]{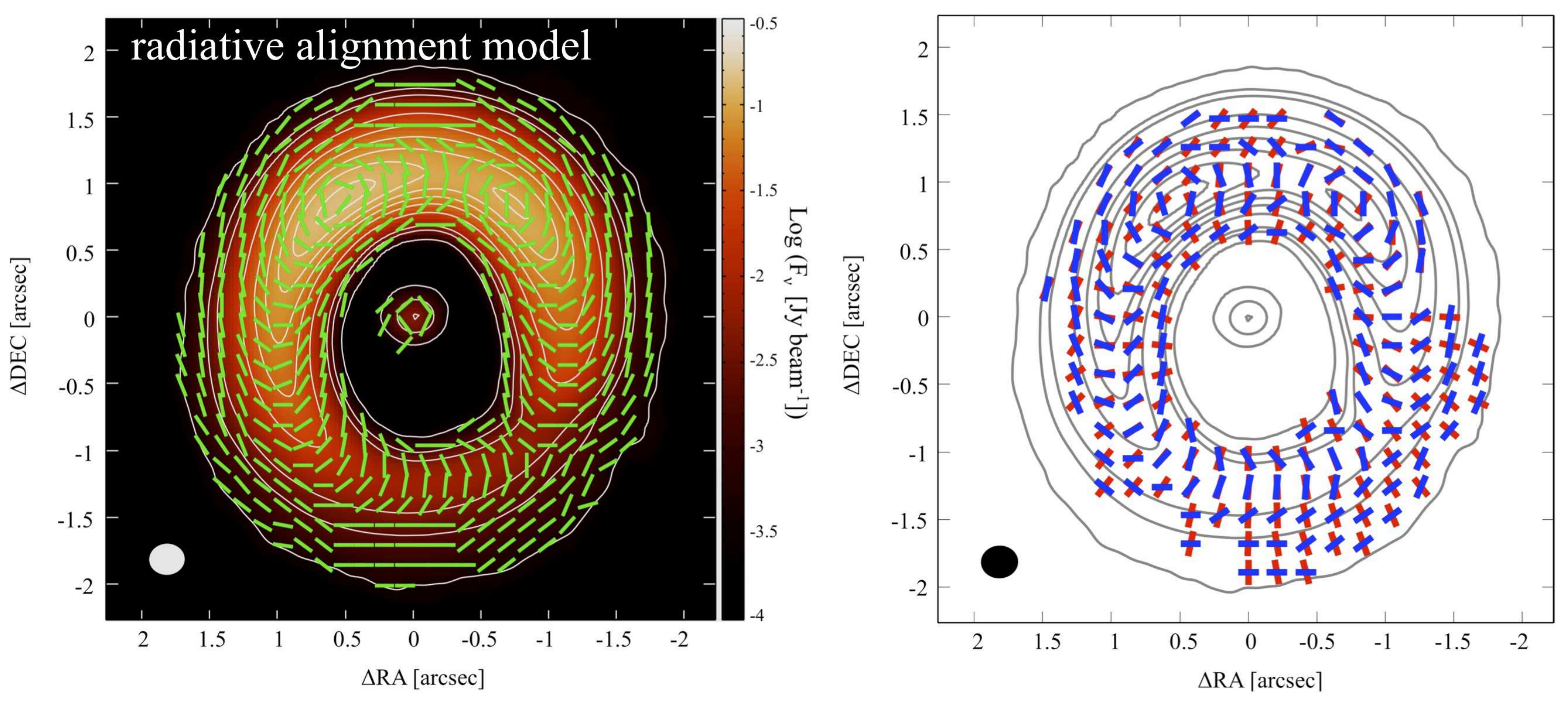}
  \end{center}
  \caption{Left panel: Polarization vectors predicted by the radiative grain alignment theory are shown in the green. The color map and the contours are the total intensity (Stokes {\it I}) map of the continuum emission. The contours are the same as those in Figure \ref{fig1}. Right panel: The polarization vectors produced by the radiative grain alignment (blue vectors) and the observations (red vectors) are plotted. The contours are the total intensity (Stokes {\it I}) map of the continuum emission same as those in the left panel.
  }
  \label{fig10}
\end{figure*}

We will now discuss polarization for radiative grain alignment in the lopsided disk of HD 142527.
To investigate the distribution of the polarization vectors, we need to know the radiation field at the midplane of the disk because the thermal dust emission at millimeter wavelengths is dominated by emission at the midplane.
\citet{taz17} showed that radiation with a wavelength of $140$ $\mu$m is efficient in aligning dust grains in the midplane.
Therefore, we need to know the radiation field with a wavelength of $\sim140$ $\mu$m.
However, there is no data with such high spatial resolution ($\sim0\farcs2$) in the $\sim140$ $\mu$m wavelength. It is also difficult to see the midplane region with $\sim140$ $\mu$m wavelength observations due to the high optical depth.
Therefore, we regard the radiation field of 0.87 mm as the radiation field of $140$ $\mu$m for simplicity.
Because both radiation wavelengths are emitted from cold thermal dust, we expect that the radiation fields of these wavelengths are similar; however, the emission of $140$ $\mu$m is optically thicker than 0.87-mm emission.

We calculate the projected intensity gradients at each pixel above the 10$\sigma$ noise level in the continuum image (Figure \ref{fig1}). 
This analysis is similar to that done by \citet{koc12}.
First, we find the pixel where $F_{\nu}\times r^{-2}$ is a maximum within the image.
Then, we calculate the gradients from the maximum pixel and define the direction to be that from the brightest pixel to the pixel of interest.
Finally, we calculate the polarization vectors by rotating the gradient directions by 90$^\circ$ because the radiative grain alignment theory states that the grains are aligned with their short axis parallel to the radiation direction and the polarization vectors are perpendicular to the radiation gradients.
Note that this calculation takes into account both inclinations of disk and dust grains because the inclined disk parallel to the dust grains is applied to the radiation field. However, it does not take into account that grains are spinning. Spinning grains may not be important for this model since the disk is almost face-on.
We show some models of the polarization vectors predicated by radiative grain alignment without the weighting of the distance in Appendix \ref{sec:Apn2}.

Figure \ref{fig10} plots our model of the polarization vectors produced by grain alignment with radiation gradients and the comparison between the model and the observations.
The model shows that the polarization vectors are parallel to the contour lines of the continuum, which is naturally expected because we assume that the polarization vectors are perpendicular to the flux gradients.
The model shows polarization vectors pointing in radial directions at the ridge of the disk and azimuthal directions in the inner and outer edge regions. 
Although the polarization vectors in the northern continuum peak regions are uncertain because of the high optical depth, the model does not match the observed polarization vectors. For example, the model predicts the polarization vectors pointing in azimuthal directions in the inner edge of the northern region of the disk, whereas the observations show the polarization vectors pointing in radial directions.
Therefore, the northern flip is unlikely due to radiative grain alignment.
We also find that the model cannot explain the radial directions of the polarization vectors in the southern region.

\subsection{Polarization Fraction}
In this section, we discuss the polarization fraction and investigate the possible mechanisms of the observed polarization. 

\begin{figure}[htbp]
  \begin{center}
  \includegraphics[width=8.5cm,bb=0 0 2584 2248]{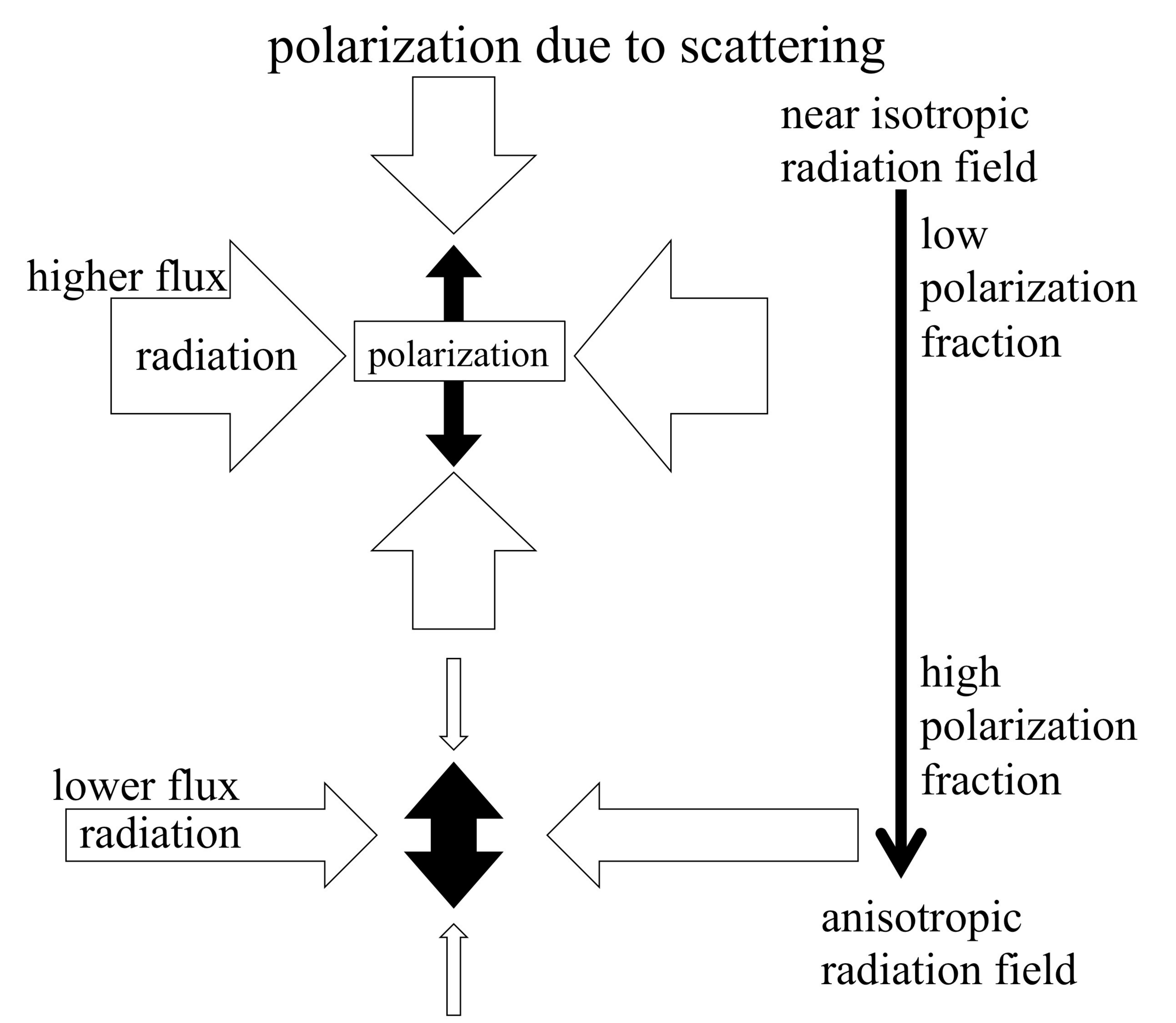}
  \end{center}
  \caption{Schematic view of polarization caused by self-scattering.
Anisotropic radiation increases the polarization fraction \citep{kat15}. }
  \label{scattering}
\end{figure}

According to the self-scattering theory, thermal dust emission can be scattered by an anisotropic radiation field of dust grains. 
The polarization fraction increases with the anisotropy of the radiation field even if the energy density of the radiation field is low (see Figure \ref{scattering}), indicating that the polarization fraction is not determined by the total energy density of the radiation field but by the degree of anisotropy of the radiation field.
Therefore, shallower (steeper) intensity variations cause lower (higher) polarization fractions.
If the intensity distribution has variations within the beam size, the polarization fraction will be reduced.
In other words, observations with a larger beam may show a lower polarization fraction due to intensity dilution.

In the case of grain alignment, the polarization fraction may not be sensitive to intensity variations, unlike the case for self-scattering.
\citet{taz17} investigate the polarization fraction for a disk with an axisymmetric intensity distribution using radiative grain alignment theory.
They found that the polarization fraction depends on the maximum grain size $a_{\rm max}$, axis ratios, shape, and optical properties of dust grains.
\citet{cho07} study grain alignment with magnetic fields within a disk and found that the polarization fraction is sensitive to the size distribution of dust grains in the disk.

To investigate the variation of the polarization fraction, we perform Gaussian fittings to the radial profiles (Figure \ref{plot1}$-$\ref{plot3}) of the total intensity and polarized intensity and examine the differences in the distributions.
For the polarized intensity distributions in the northern region (Figures \ref{plot1} and \ref{plot2}), we only fit the data outside the local peaks where the radius is $r\lesssim120$ au or $r\gtrsim230$ au in order to avoid the optical depth effect.

The full width half maximum (FWHM) of the total intensity and polarized intensity profiles are $73\pm1$ au and $64\pm2$ au in Figure \ref{plot1}, $76\pm1$ au and $68\pm2$ au in Figure \ref{plot2}, and $76\pm1$ au and $74\pm1$ au in Figure \ref{plot3}, respectively.

These results indicate that the distributions of the total intensity are shallower than those of the polarized intensity in the northern region. 
Using observations with different spatial resolutions, the polarization fractions derived from synthesized images will also be different because of the difference in radial profiles between the total and polarized intensities. 
In contrast, the FWHMs for the southern region profiles of the total intensity and polarized intensity are almost the same.
Therefore, the polarization fraction does not depend on the spatial resolution.

Figure \ref{fig6} shows that the polarization fraction depends on the spatial resolution and that there is a large variation of the ratio in the northern region.
In the self-scattering theory, high ratios at the inner edge indicate steep intensity gradients; these gradients may not have been resolved by our observations.
The grain alignment with magnetic fields and radiation gradients theories do not predict variations of the polarization fraction on beam sizes assuming the curvature is sufficiently resolved. Therefore, the self-scattering may be preferred at least in the inner edge of the north part of the disk.
However, it is possible that the polarization fraction increases with increasing spatial resolution if the polarization has non-uniform patterns.
Therefore, we cannot rule out the grain alignment theories.

In contrast, the polarization fraction ratio for the southern region is almost unity and does not depend on the spatial resolution.
The constant value of the polarization fraction on the spatial resolution may indicate that self-scattering is unlikely in the southern region or that the previous observations already resolved the intensity distribution in the southern region.

As shown in Figure \ref{fig9}, we derive the polarization fraction in the southern region to be as high as $\sim15$\%, which is consistent with the value  ($\gtrsim10$\%) predicted with grain alignment with magnetic fields in a resolved disk \citep{cho07}.
\citet{ber17} recently performed three-dimensional radiative transfer simulations of polarized emission from dust grains aligned with magnetic fields. They also reported a polarization fraction of $\sim10$\% in toroidal magnetic fields. Therefore, grain alignment with magnetic fields in the southern region of the disk is a reasonable explanation for the observed polarization.
In contrast, such a high polarization fraction is not expected with self-scattering since the polarization degree is determined by the degree of anisotropy of the radiation field. A maximum polarization fraction of $\sim3$\% was calculated by \citet{kat16} with the assumption of density contrast for the HD 142527 disk.
However, a polarization fraction in self-scattering may increase with increasing the axis ratio of dust grains \citep[][]{yan16b}.

\subsection{Polarization in HD 142527}

We discussed the possible mechanisms for the polarization in the view of the morphology of the polarization vectors in Section 4.1 and the polarization fraction in Section 4.2. In this section, we consider what causes the polarization in this disk and what kind of dust distribution model is necessary.

\begin{table*}
\centering
\caption{\label{table}Possible mechanisms for polarization}
\begin{tabular}{p{70mm}lcc}\hline\hline
Polarization & Northern region & Southern region \\\hline
Self-scattering & likely	& unlikely (?)	\\
Grain alignment with magnetic fields & unlikely	& likely	\\
Grain alignment with radiation gradients &unlikely	&	unlikely\\\hline
\end{tabular}
\end{table*}

We summarize the possibilities of the three polarization mechanisms in the northern and southern regions in Table \ref{table}.
It is likely that the self-scattering theory is consistent with the observations in the northern region because the flip of the polarization vectors sharply occurs outside the northern disk. In contrast, grain alignment with magnetic fields is more likely in the southern region because no flip is found there. Toroidal magnetic fields are likely to explain the radial directions of the polarization vectors, at least in the southern region.
Self-scattering is less likely but cannot be fully ruled out because it might be possible that a flip is detected outside of the disk in the southern region with high sensitivity observations and because the high polarization fraction may be explained by the axis ratio of dust grains.
Radiative grain alignment is unlikely either in the northern or southern region as the model of radiative grain alignment does not fit well with the observations, as shown in Section 4.1.

To explain the difference in the polarization mechanism between the northern and southern regions, we discuss the differences in the dust size distribution between these two regions.
In general, lopsided disks are explained by dust concentration \citep[e.g.,][]{bir13}; this applies to the disk of HD 142527 \citep{mut15,boe17}.
A gas pressure bump traps dust particles \citep[][]{pin12,bir13,lyr13}.
In particular, \citet{bir13} show that the largest grains ($\sim$ cm) are efficiently trapped near the pressure maximum.
The grain size is possibly as large as a centimeter in the northern region and smaller in the southern region.

The lopsided disk suggests that the grain size is on the scale of a centimeter, whereas polarization due to self-scattering at a 0.87-mm wavelength indicates that the grain size should be $\sim150$ $\mu$m \citep{kat15}. To understand this difference in size estimations, we consider the vertical structures.
Taking into account ALMA studies that showed the northern region is optically thick ($\tau\gtrsim2$) \citep{fuk13,cas15}, the polarized emission may come from a relatively upper layer of the disk.
It is possible that the photosphere (a region where $\tau=1$) is located at a layer with a dust grain size of $\sim150$ $\mu$m and that larger grains settle in the midplane \citep{dur95,you07}.
In this case, the grain size may be larger than 150 $\mu$m at the midplane.

As another explanation, the fluffy structure of dust including porosity may explain the scattering efficiency inferred from the polarization observations as well as the flat spectral index \citep{kat16}.
Fluffy dust with a filling factor of $10^{-4}$ has large scattering opacity \citep{kat13,kat14,taz16}, and its scattering efficiency is much higher than that for compact dust \citep{kat16}.
In this case, grains with porous and massive aggregates dominate the thermal dust emission, and settle in the midplane.
They may account for the scattering at millimeter wavelengths even though the aggregate radius is much larger than the wavelength.
Then, we may detect polarization due to self-scattering at a 0.87-mm wavelength.

\begin{figure}[htbp]
  \begin{center}
  \includegraphics[width=8.5cm,bb=0 0 2895 2032]{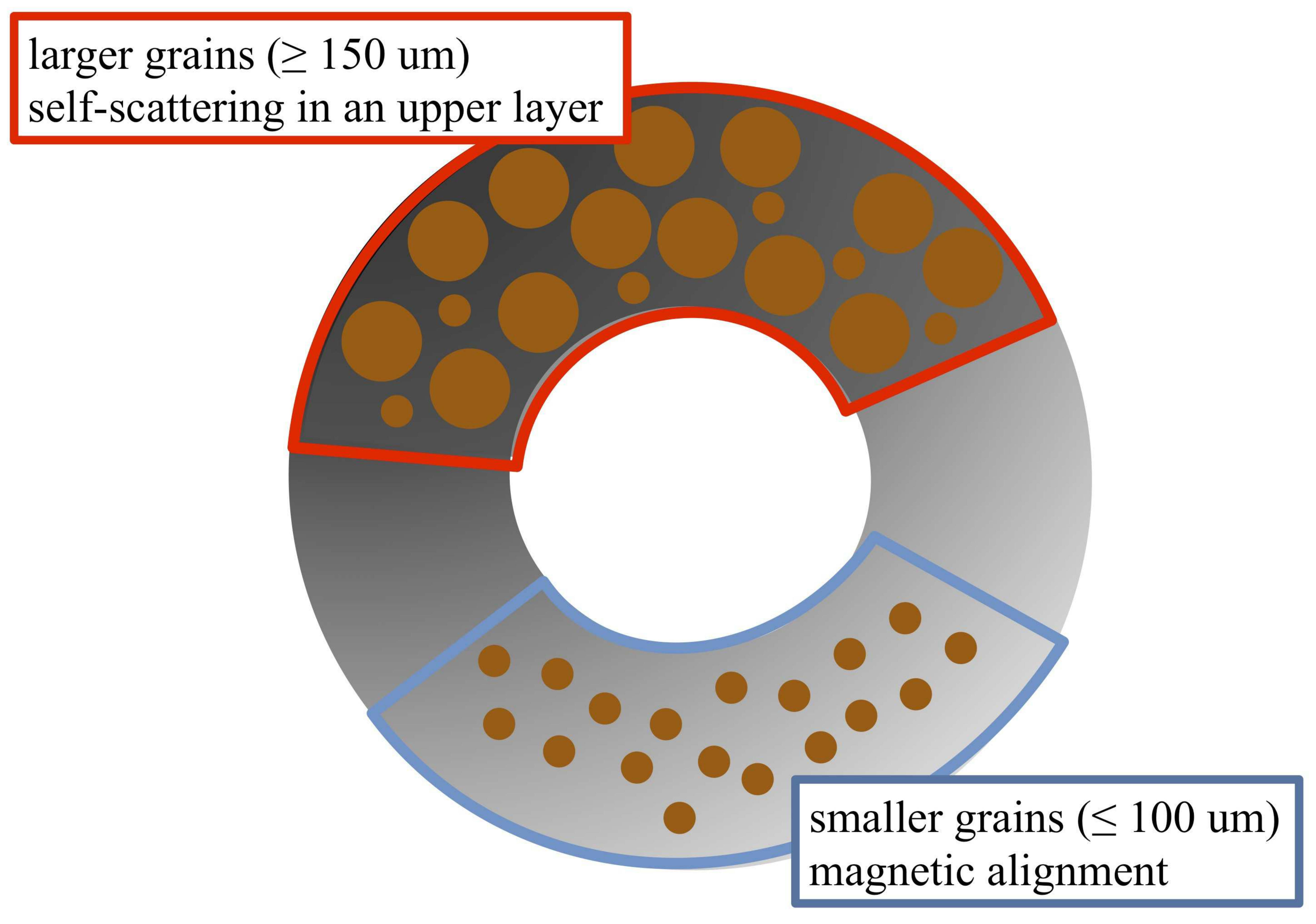}
    \end{center}
  \caption{Schematic diagram of the disk around HD 142527. The gray-scale intensity represents the lopsided intensity distribution of the disk.
  The regions outlined in red and blue are the polarized regions due to self-scattering and grain alignment with toroidal magnetic fields, respectively. These different polarization mechanisms can be explained by the different grain size distributions.
  }
  \label{hd142527view}
\end{figure}

In contrast, the thermal dust emission in the southern region, where the polarization is consistent with grain alignment with magnetic fields, may be dominated by small dust grains.
\citet{taz17} show that small grains, around 10 microns in size, can align with magnetic fields if they have supermagnetic inclusions and/or the magnetic fields are strong.
Small dust grains produce almost no polarization due to self-scattering.
Therefore, the polarization in the southern region may be due to small grains (at least $\lesssim100$ $\mu$m) aligned with magnetic fields.
A schematic diagram of the dust distributions is shown in Figure \ref{hd142527view}.

\begin{figure}[htbp]
  \begin{center}
  \includegraphics[width=8.5cm,bb=0 0 2380 2212]{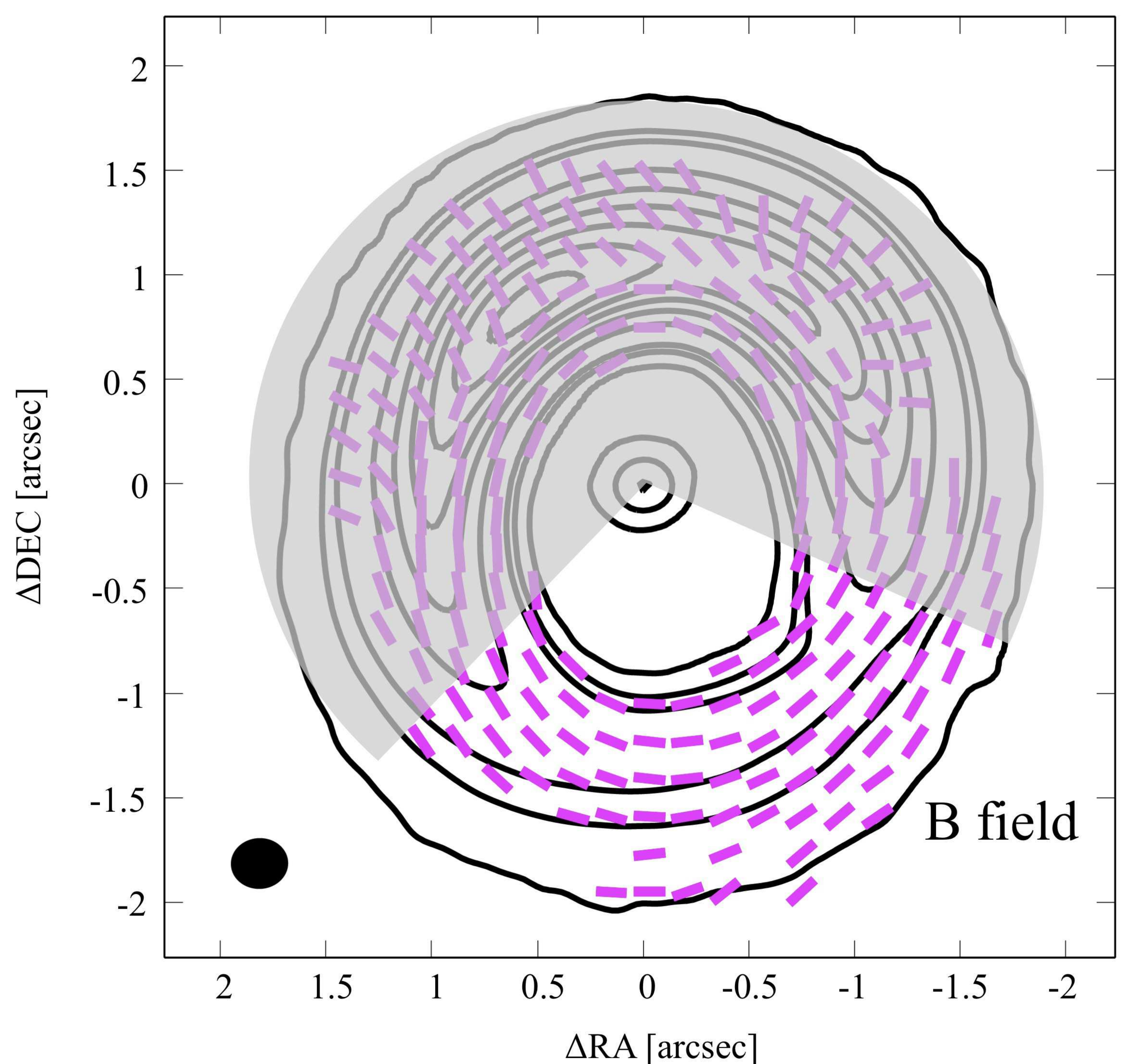}
    \end{center}
  \caption{Line segments represent the magnetic field orientations, rotated by $90^\circ$ from dust polarization. The contours are the same as those in Figure \ref{fig1}. The northern region does not show magnetic fields because the polarization there is not produced by grain alignment with magnetic fields.
  }
  \label{mag}
\end{figure}

The toroidal magnetic fields are likely to explain the southern distribution of the polarization vectors as described in Section 4.1.
We plot the magnetic fields distribution in Figure \ref{mag} by rotating the polarization vectors by $90^\circ$.
The magnetic fields are toroidal in the southern part.
The northern region does not show magnetic fields because the polarization there is not produced by grain alignment with magnetic fields.

\citet{mut15} showed that the density contrast between the northern and southern regions is $\sim70$ assuming a maximum dust grain size of 1 mm.
\citet{soo17} derived the contrast of about 40 by reducing the dust scattering opacity in the northern region to fit the intensity profile.
However, if the dust grain size is as small as a micron in the southern region, the dust opacity should be lower, indicating that these surface density contrasts were overestimated.
\citet{ric10} reported that the dust opacity varies by a factor of at most $\sim5$ for a dust grain size in sub-micron to millimeter range.
Therefore, the density contrast would be 5 times lower than $40-70$ assuming a maximum dust grain size of $\sim$ microns in the southern region.
Thus, the intensity contrast between the northern and southern regions cannot be explained by only the dust opacity variation with grain size. The surface density contrast is still needed to explain the horseshoe-like structure of the continuum emission.

\subsection{Future Prospects}
Our results demonstrate that magnetic fields can be found based on millimeter polarization even in protoplanetary disks if thermal emission is dominated by small dust grains.
However, it is very important to estimate the dust grain size properly, and investigate further the condition of grain alignment with magnetic fields using theoretical/numerical calculations.
In the future, we distinguish the polarization caused by scattering from that caused by grain alignment by observing multiple millimeter wavelengths, because self-scattering depends on the observation wavelength.
\citet{alv18} have investigated multi-frequency polarization of a circumbinary disk around a Class I source with  ALMA Band 3, 6, and 7 observations. They have shown that  polarization patterns remarkably consistent among all three bands and have indicated that polarization in this source is produced by magnetic fields.
According to these results, we predict that the polarization will not significantly change in the southern region for other millimeter wavelengths.
We also predict that the polarization vectors will probably change in the northern region with longer-wavelength (e.g., 3 mm) polarization observations because radiative grain alignment will be efficient at such wavelengths.
If the 3-mm continuum emission is optically thin and the maximum grain size is as large as 500 $\mu$m in the northern region, the polarization will be caused by a combination of self-scattering and radiative grain alignment.

The possibility of mechanical grain alignment also needs to be discussed. In this paper, we do not investigate this mechanism because no polarization model for mechanical grain alignment has been studied in protoplanetary disks. Further studies of the mechanical grain alignment theory are needed.

An inclined disk has been inferred in the inner disk \citep{mar15}.
We expect that self-scattered-polarized emission from the inner disk will be detected because the inclined disk produces an anisotropic radiation field.
Therefore, we may be able to measure the inclination of the inner disk from the distributions of the polarization vectors with higher sensitivity and spatial resolution.

\section{Conclusion}

The polarization mechanisms at millimeter wavelengths have been proposed: (1) grain alignment with magnetic fields, (2) grain alignment with radiation gradients, and (3) self-scattering of thermal dust emission.
Aiming to identify these mechanisms in the protoplanetary disks, we presented ALMA polarization observations of the 0.87-mm dust continuum toward the circumstellar disk around HD 142527.
Our ALMA polarization observations were performed with a resolution of  $0\farcs27\times0\farcs24$, corresponding to a spatial resolution of $\sim38\times34$ au, assuming a distance of 156 pc (Gaia Collaboration).
Our main results are summarized as follows.

1. We found that the polarized intensity exhibits a ring-like structure and that the polarization vectors mainly point in radial directions on this ring. 
Polarization vectors pointing in azimuthal directions were found outside the northern region. This flip of the polarization vectors from radial to azimuthal directions can be explained by the self-scattering of thermal dust emission, as shown by \citet{kat16}.

2. We compared our high spatial resolution data with the previous data reported by \citet{kat16}. The Stokes {\it I}, {\it Q}, and {\it U} images based on the two observations looked similar.
However, our high spatial resolution data identified the region where the polarization intensity drops and the flip occurs more clearly. We found that the polarization fraction increases with increasing spatial resolution in the inner edge of the northern region of the disk.

3. We detected an inner circumbinary disk in the Stokes {\it I} emission. 
We did not detect polarization, indicating an upper limit of the polarization fraction of $\sim2.0$\% from the 3$\sigma$ noise level.

4. We discussed whether the polarization in the disk is due to grain alignment or self-scattering by investigating the morphology of the polarization vectors. We found that the flip of the polarization vectors in the northern region can be explained by self-scattering, while the radial distributions of the polarization vectors in the southern region are likely explained by grain alignment with toroidal magnetic fields.
Self-scattering is less likely but cannot be fully ruled out in the southern region.
We constructed the simple model of the polarization vectors for radiative grain alignment. However, the model does not fit well with the observations.

5. We found the polarization fraction in the southern region to be as high as $\sim15$\%. This is consistent with the value ($\gtrsim10$\%) predicted by grain alignment with magnetic fields. In contrast, self-scattering is difficult to explain such a high polarization fraction since polarization degree is determined by the degree of the anisotropy of the radiation field. However, it may be possible that a polarization fraction in self-scattering increases with increasing the axis ratio of dust grains.

6. To explain the different polarization mechanisms between the northern and southern region, we discussed the differences of the dust grain size.
Small dust grains ($\lesssim$ 100 micron) are dominant and aligned with the magnetic fields in the southern region, while middle-sized ($\sim100$ micron) grains in the upper layer emit the self-scattered polarized emission in the northern region. The grain size near the middle plane in the northern region cannot be measured because the emission at 0.87 mm is optically thick. However, it can be speculated that larger dust grains ($\gtrsim$ cm) may accumulate near the middle plane due to the dust trap.
We showed that the magnetic fields are toroidal at least in the southern region.
\\

As future prospects, we will distinguish the polarization caused by self-scattering from that caused by grain alignment by observing multiple millimeter wavelengths because self-scattering depends on the observation wavelength.
We predict that polarization will not significantly change in the southern region for other millimeter wavelengths and that it will probably change in the northern region.

\acknowledgments
We gratefully appreciate the comments from the anonymous referee that significantly improved this article.
S.O. thanks Ryo Tazaki and Nami Sakai for helpful discussions.
This work is supported by Japan Society for Promotion of Science (JSPS) KAKENHI (No. 18K13595).
This paper makes use of the following ALMA data: ADS/JAO.ALMA\#2015.1.01025.S. ALMA 
is a partnership of ESO (representing its member states), NSF (USA) and NINS (Japan), 
together with NRC (Canada), NSC and ASIAA (Taiwan), and KASI (Republic of Korea), in 
cooperation with the Republic of Chile. The Joint ALMA Observatory is operated by 
ESO, AUI/NRAO and NAOJ.

Data analysis was in part carried out on common use data analysis computer system at the Astronomy Data Center, ADC, of the National Astronomical Observatory of Japan.



\facility{ALMA}

\software{CASA \citep[v4.5.3; ][]{mcm07}}

\appendix\section{High Spatial Resolution Image and Spiral Arms}\label{sec:Apn1}

\begin{figure}[htbp]
  \begin{center}
  \includegraphics[width=16cm,bb=0 0 2680 2201]{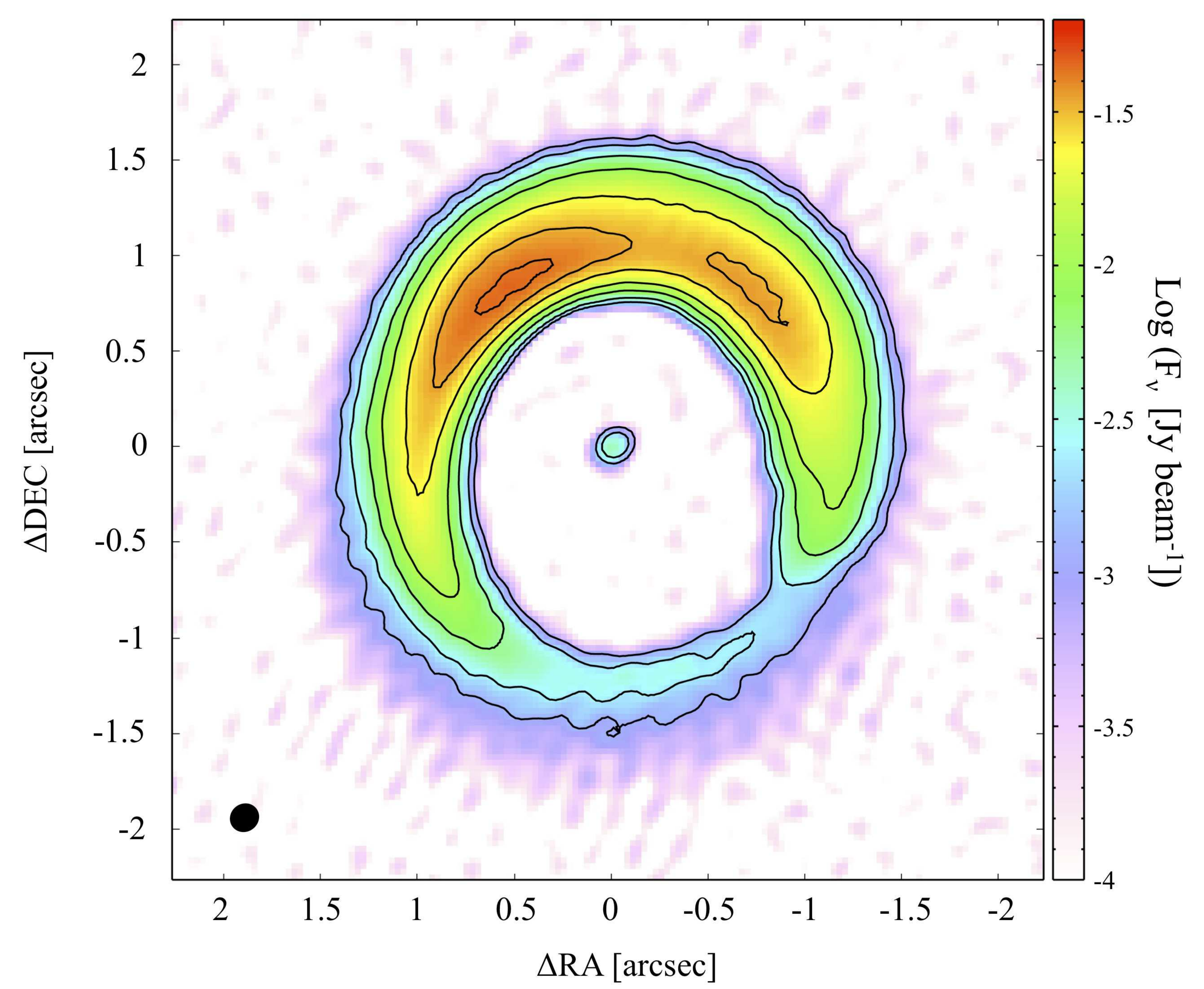}
  \end{center}
  \caption{Total intensity (Stokes {\it I}) map of the continuum emission at 0.87 mm with Briggs weighting with a robust parameter of $-2$. The synthesized beam, with a size of $0\farcs15\times0\farcs13$ and a position angle of P.A. $=-52.0^\circ$, is indicated by the filled ellipse in the bottom left corner.
  The contours correspond to $(10,20,50,100,200,300,380)\times\sigma_I$, where $\sigma_I$ is $1.14\times10^{-4}$ Jy beam$^{-1}$.
  }
  \label{fig7}
\end{figure}

Our polarization observations were done with high sensitivity to detect polarized emission.
We create a high spatial resolution image for the total intensity map using Briggs weighting with a robust parameter of $-2$.
Figure \ref{fig7} shows the total intensity with the highest spatial resolution.

\begin{figure*}[htbp]
  \begin{center}
  \includegraphics[width=16cm,bb=0 0 2866 1324]{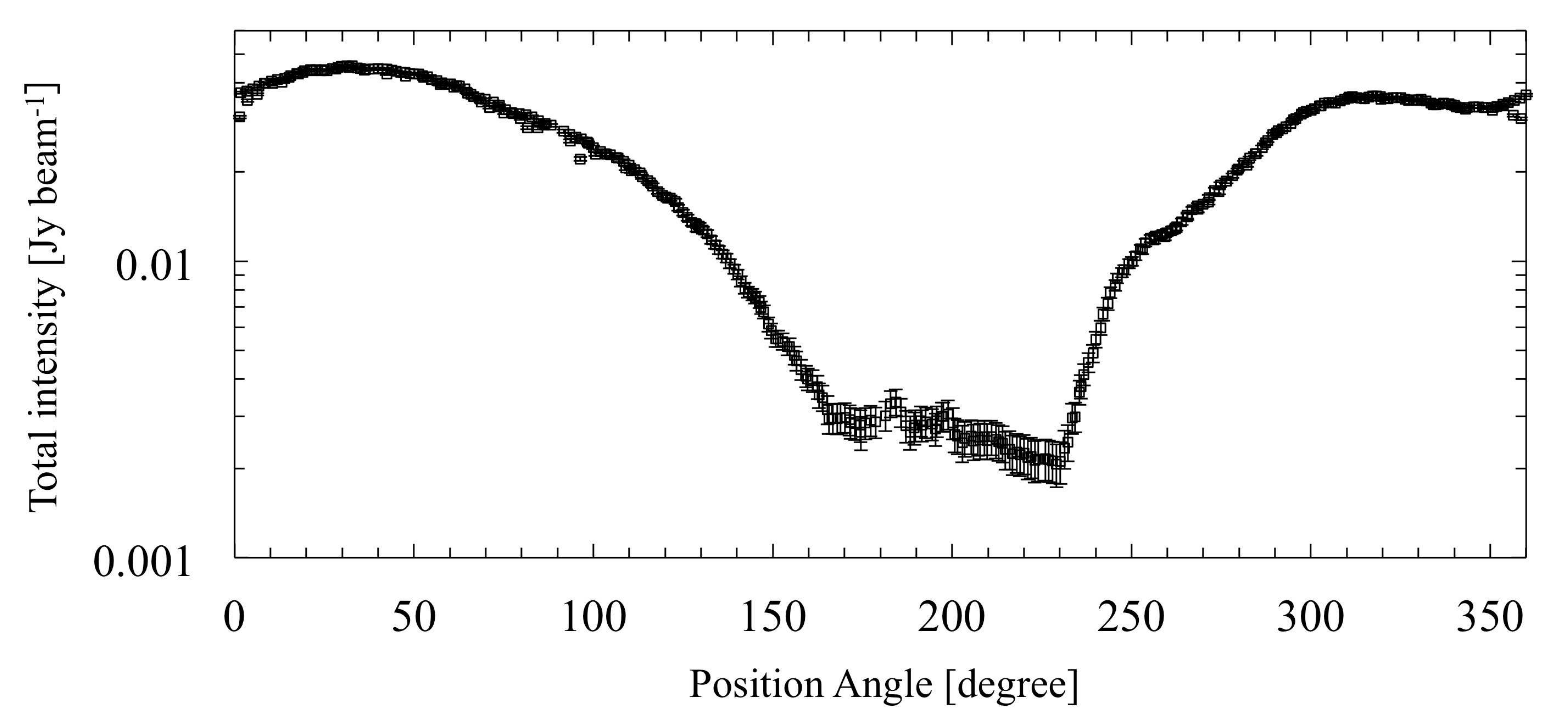}
  \end{center}
  \caption{Peak intensity on horseshoe structure as a function of position angle.
  }
  \label{plot7}
\end{figure*}

\begin{figure}[htbp]
  \begin{center}
  \includegraphics[width=16cm,bb=0 0 2493 2224]{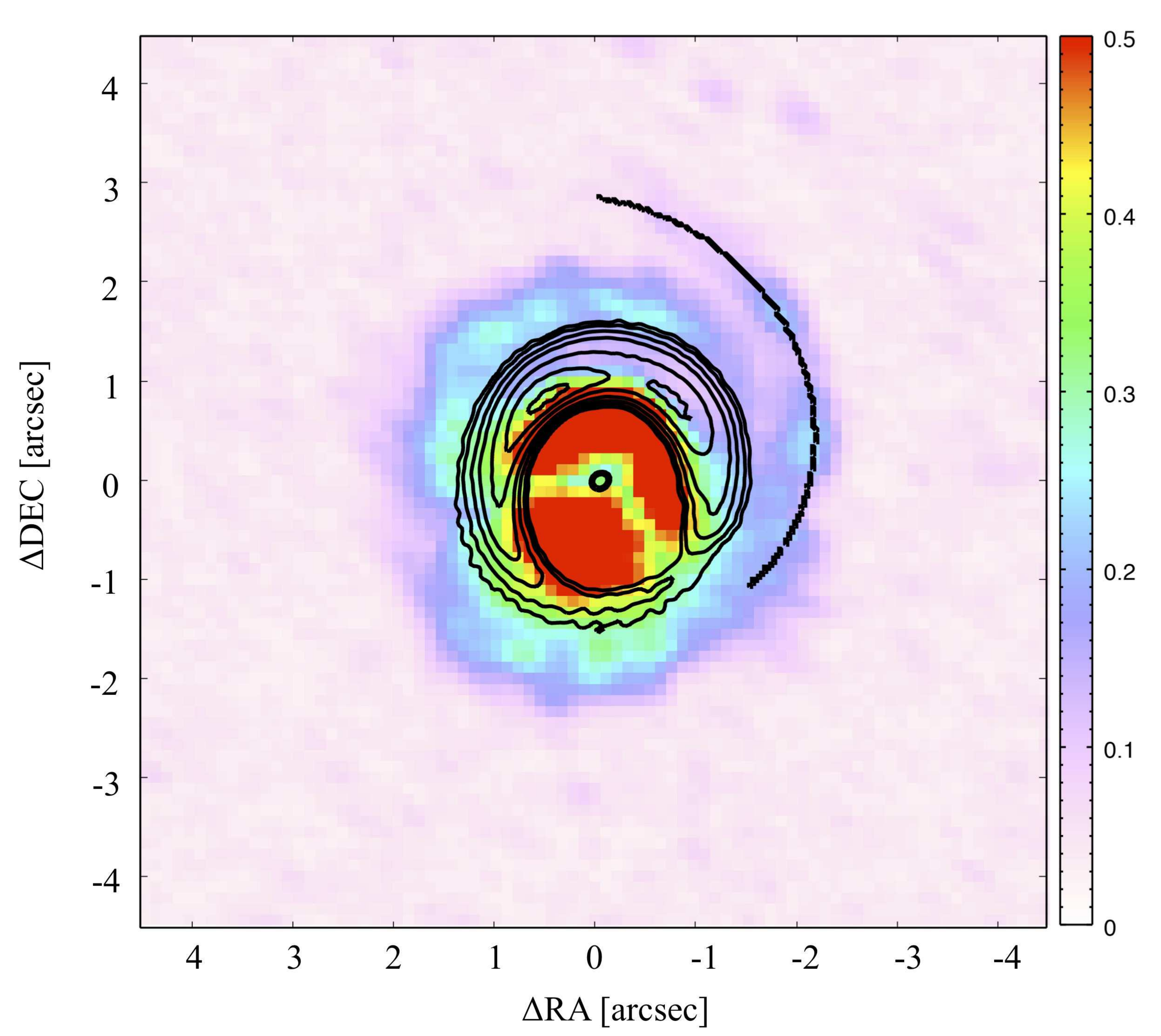}
  \end{center}
  \caption{Color map of the CO $J=3-2$ peak intensity overlaid with the innermost spiral arm from \citet{chr14}. The contours are the same as those in Figure \ref{fig7}
  }
  \label{fig8}
\end{figure}

The beam size is $0\farcs15\times0\farcs13$, corresponding to a spatial resolution of $\sim21\times18$ au.
The contours are $(10,20,50,100,200,300,380)\times\sigma_I$, where $\sigma_I$ is $1.143\times10^{-4}$.
The total flux is about 3.0 Jy and the peak intensity is 46 mJy beam$^{-1}$ in the northeast region.
The minimum intensity of the horseshoe structure at P.A. $=230^\circ$ is 2.1 mJy beam$^{-1}$.
The contrast in the intensity reaches a ratio of $22:1$ between the northeast peak and the southwest minimum, which is consistent with the results reported by \citet{mut15,boe17}.
At the position of the star, emission was detected with an intensity of 4.3 mJy beam$^{-1}$.
Gaussian fitting resulted in an FWHM of $0\farcs15\times0\farcs13$, indicating that the inner disk is still spatially unresolved.
Figure \ref{plot7} shows the peak intensity on the horseshoe structure as a function of the position angle.
The weak intensity is flatly distributed from P.A. $=165^\circ$ to P.A. $=230^\circ$.
The minimum intensity is found at P.A. $=230^\circ$: the intensity increases rapidly with increasing position angle.
The profile seems to change at P.A. $=165^\circ$ and P.A. $=230^\circ$, which might represent changes in the dust distribution at these positions.

\citet{chr14} identified spiral arms from ALMA observations of $^{12}$CO $J=2-1$, $^{12}$CO $J=3-2$, and $^{13}$CO $J=2-1$ lines.
The innermost spiral arm seems to continue to the outer disk.
Therefore, we investigate the connection between the innermost spiral arm and the disk using ALMA archival data.
Figure \ref{fig8} shows the peak intensity map for the CO $J=3-2$ line overlaid with the continuum emission.
The black spiral represents the innermost spiral identified by \citet{chr14}.
A similar spiral arm was reported from near-infrared (NIR) observations \citep[e.g.,][]{fuk06,2012ApJ...754L..31C,2012A&A...546A..24R,2013A&A...556A.123C,2014ApJ...781...87A} and the CO spiral arm has been suggested to be part of the same structure.
Because the spiral arm continues to P.A. $=180^\circ$ in the NIR spectrum, the spiral arm is likely to be connected to the southern region of the disk.
Therefore, the features of the lopsided intensity distribution may be affected by the dynamic motion of this spiral arm.

\section{Polarization Vectors with the Radiative Grain Alignment}\label{sec:Apn2}

\begin{figure}[htbp]
  \begin{center}
  \includegraphics[width=16cm,bb=0 0 2690 2248]{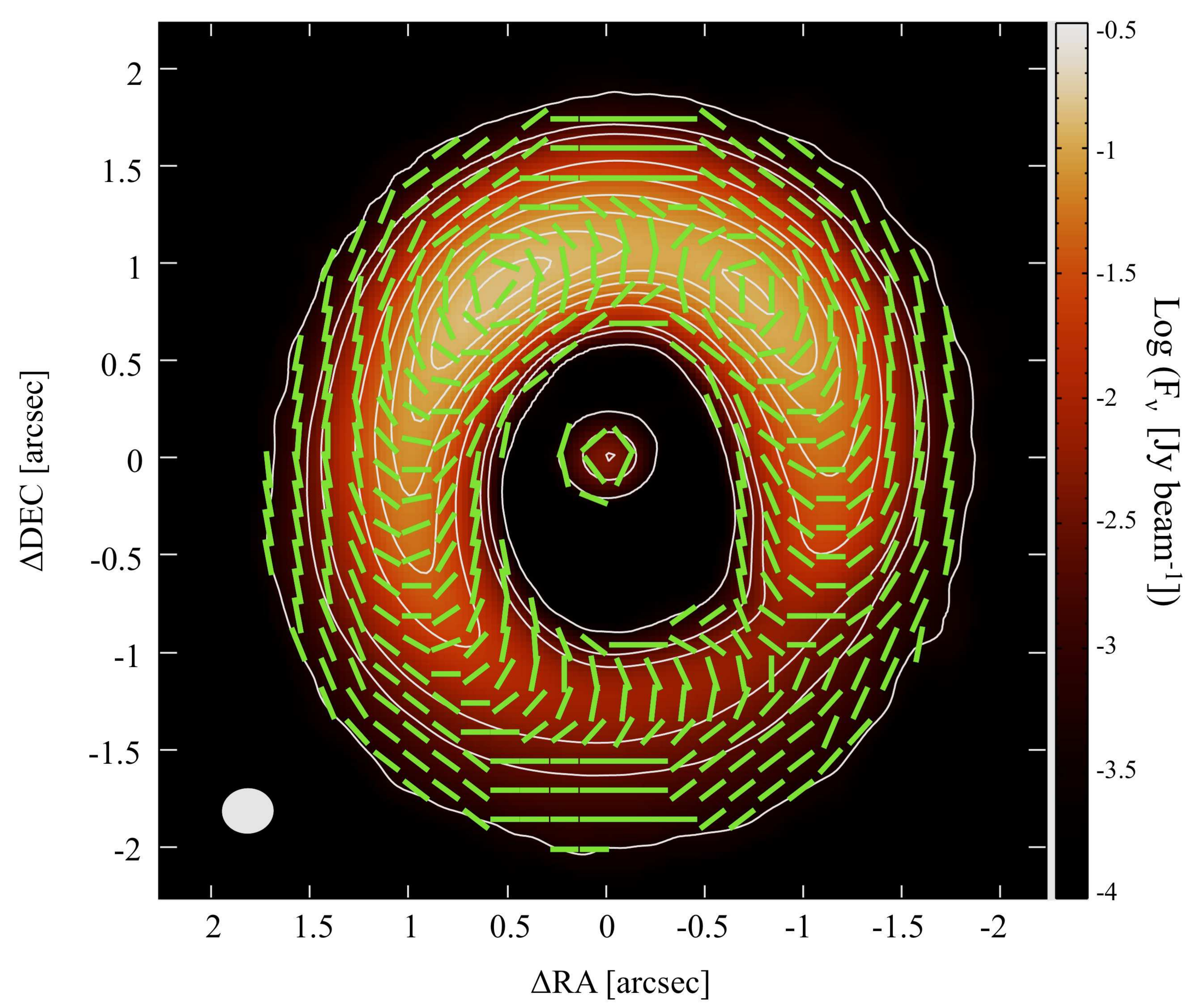}
    \end{center}
  \caption{Polarization vectors predicted by the radiative grain alignment theory without the weighting of the distance. A searching radius where the radiation is taken into account, is $1\times$ beam size.
  }
  \label{beam}
\end{figure}

\begin{figure}[htbp]
  \begin{center}
  \includegraphics[width=16cm,bb=0 0 2673 2248]{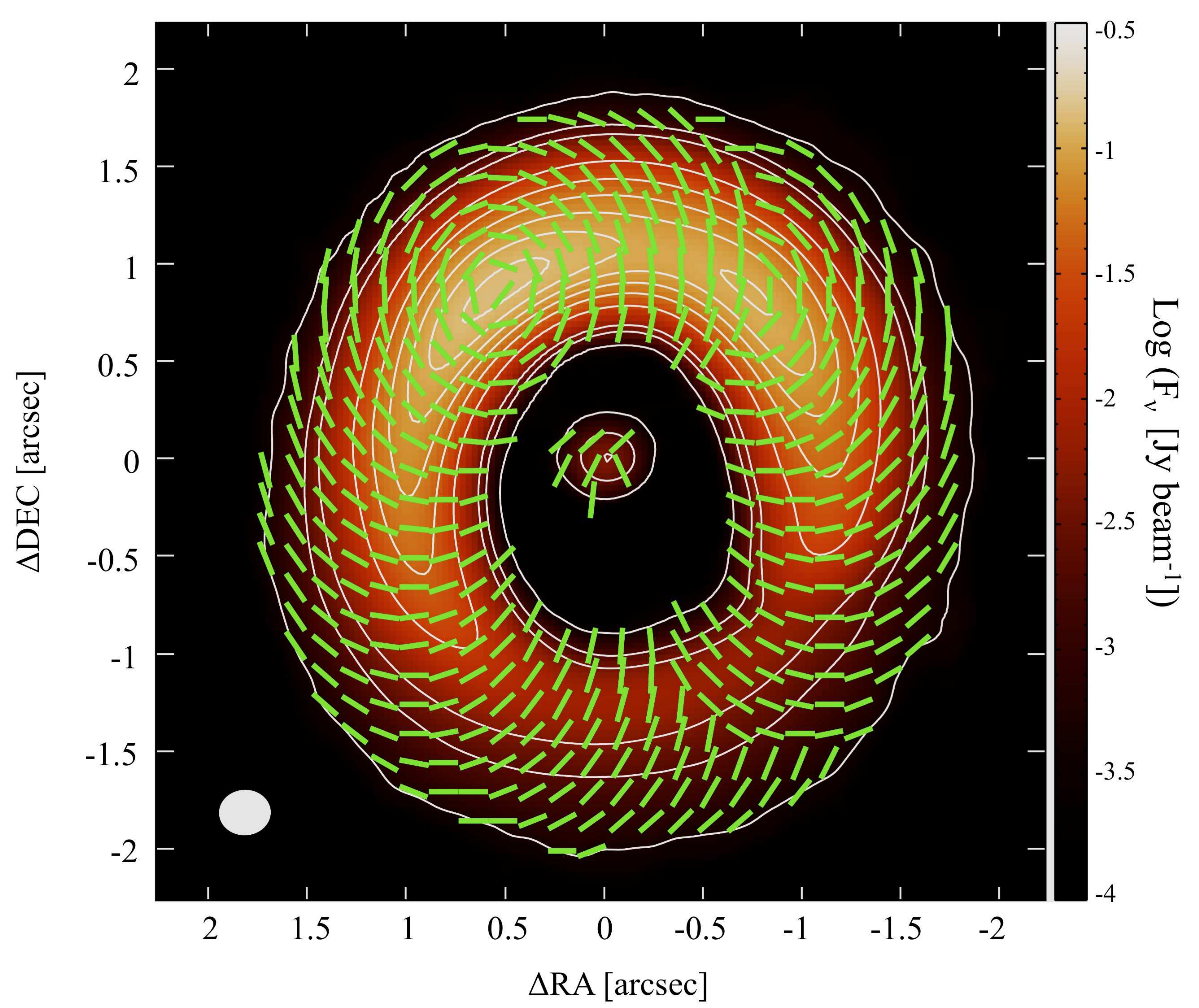}
    \end{center}
  \caption{Same as Figure \ref{beam} but with a searching radius of $3\times$ beam size.
  }
  \label{3beam}
\end{figure}

\begin{figure}[htbp]
  \begin{center}
  \includegraphics[width=16cm,bb=0 0 2681 2248]{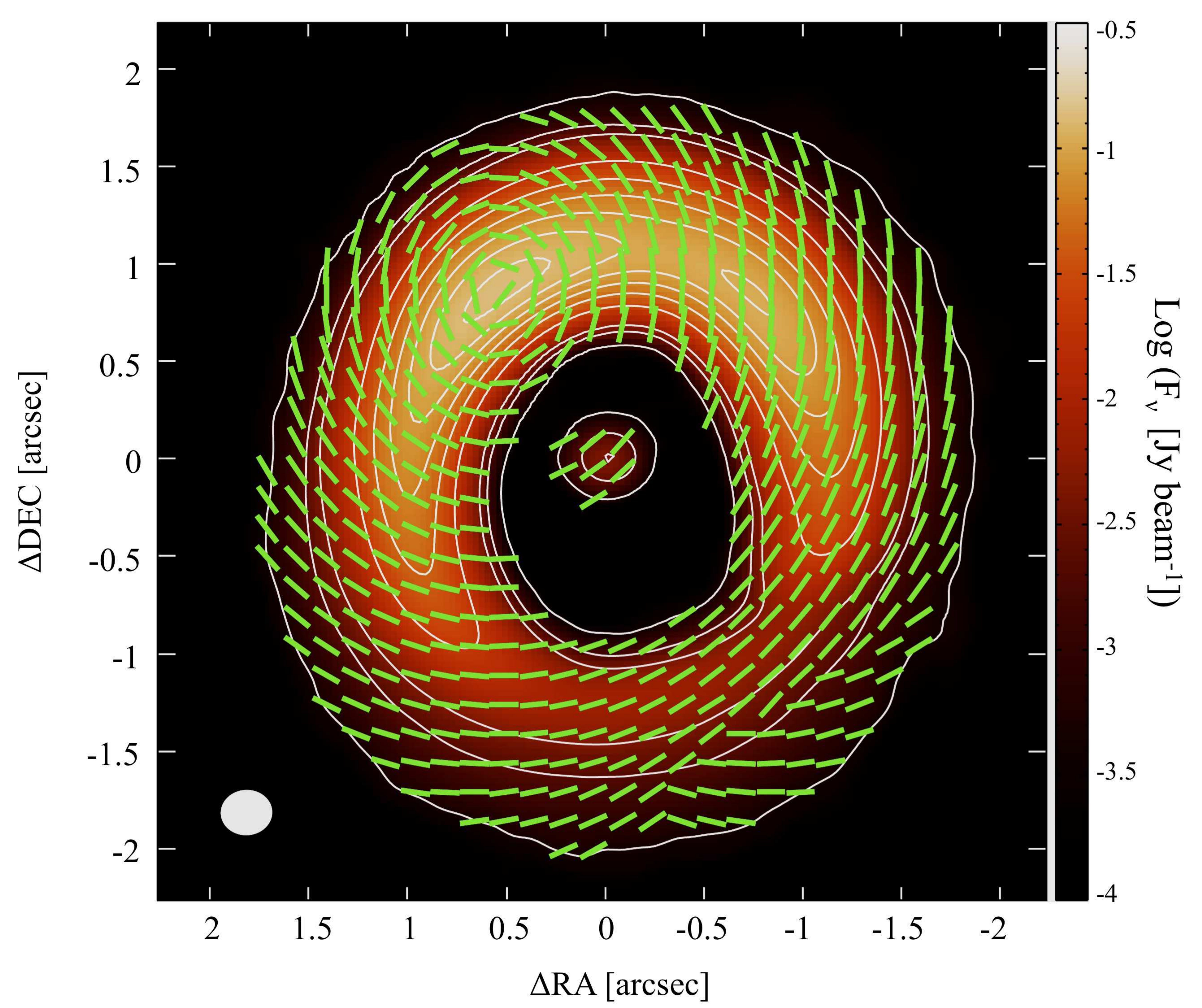}
    \end{center}
  \caption{Same as Figure \ref{beam} but with a searching radius of $10\times$ beam size.
  }
  \label{10beam}
\end{figure}

In Section 4.2, we show the polarization vectors predicted by radiative grain alignment (see Figure \ref{fig10}).
The radiation flux is weighted by the square of the distance because flux decreases with $r^{-2}$.
In this appendix, we show the polarization vectors predicted by radiative grain alignment without the weighting of the distance. Figures \ref{beam} $-$ \ref{10beam} show radiative grain alignment for three searching radii: $1\times$ beam size, $3\times$ beam size, and $10\times$ beam size. The searching radius is the region where the radiation is taken into account for calculating the gradient. These figures allow us to find the dominant radiation area for the polarization with radiative grain alignment.
Figure \ref{beam}, with a searching radius of $1\times$ beam size, is mostly similar to Figure \ref{fig10}, indicating that the nearby radiation affects the polarization. 
Figure \ref{3beam}, with a searching radius of $3\times$ beam size, shows azimuthal directions around the dust local peaks.
Figure \ref{10beam}, with a searching radius of $10\times$ beam size, shows the azimuthal direction around the dust peak position because the radiation mainly comes from the peak intensity position without the weighting of the distance.

\end{document}